\documentclass[a4paper,11pt]{article}
\pdfoutput=1 % if your are submitting a pdflatex (i.e. if you have
\usepackage{jheppub} % for details on the use of the package, please
                     \usepackage{graphicx}

\newcommand{\arXiv}[2]{\href{http://arxiv.org/abs/#1}{{\tt arXiv:#2}}}
\newcommand{\hep}[2]{\href{http://arxiv.org/abs/#1}{{\tt #2}}}

\newcommand{\rg}{\mathrm{g}}
\newcommand{\rK}{\mathrm{K}}
\newcommand{\rL}{\mathrm{L}}
\newcommand{\rR}{\mathrm{R}}
\newcommand{\rT}{\mathrm{T}}
\newcommand{\rU}{\mathrm{U}}
\newcommand{\rV}{\mathrm{V}}

\def\defeq{\stackrel{\text{def}}=}
\def\({\left(} \def\){\right)}
\def\hf{ {\textstyle{1\over 2}} } \def\hfi{ {\textstyle{i\over 2}} }

 \def\hfi{ {\textstyle{i\over 2}} }
   \def\CO{{
\mathcal{ O} }}  
 
 \def\CC{ {\mathcal C}}
   \def\CN{{ \cal N}}  
     \def\a{\alpha} \def\b{\beta} 
\def\e{\epsilon}   
\def\th{\theta}   
   \def\xx{
{ { \bf x} }}

\def\yy{ { { \bf y} }}
 
\def\uu{ { {\bf u} }}  \def\ww{ { \bf
w }} \def\thth{ { \bm {\th } }}

                    \def\II{{\mathbb{I}}}
                     
                   \def\so{\varepsilon}
\def\fn#1{\footnote{#1}}
\def\bm#1{\text{\boldmath $#1$}}
\def\ket#1{|#1\rangle}
\def\bra#1{\langle #1 |}
\def\period{\,.}
\newcommand\encadremath[1]{\vbox{\hrule\hbox{\vrule\kern8pt
\vbox{\kern8pt \hbox{$\displaystyle #1$}\kern8pt}
\kern8pt\vrule}\hrule}} \def\enca#1{\vbox{\hrule\hbox{
\vrule\kern8pt\vbox{\kern8pt \hbox{$\displaystyle #1$} \kern8pt}
\kern8pt\vrule}\hrule}}
\def\LL{ L}

\title{The hexagon in the mirror: the three-point function in the SoV
representation}

 \author{Yunfeng Jiang$^a$,}
\author{Shota Komatsu$^b$,}
\author{Ivan Kostov$^a$,}
\author{Didina Serban$^a$}

\emailAdd{yunfeng.jiang; ivan.kostov; didina.serban@cea.fr; skomatsu@perimeterinstitute.ca}

\affiliation{$^a$ Institut de Physique Th\'eorique,
DSM, CEA, URA2306 CNRS\\Saclay, F-91191 Gif-sur-Yvette,
France}
\affiliation{$^b$ Perimeter Institute for Theoretical Physics, Waterloo, Ontario, Canada}

\abstract{ We derive an integral expression for the leading-order
type I-I-I three-point functions in the $\mathfrak{su}(2) $-sector of
$\mathcal{N}=4$ super Yang-Mills theory, for which no determinant
formula is known. To this end, we first map
the problem to the partition function of the six vertex model with a
hexagonal boundary.  The advantage of the six-vertex model expression
is that it reveals an extra symmetry of the problem, which is the
invariance under 90$^{\circ}$ rotation.  On the spin-chain side, this
corresponds to the exchange of the quantum space and the auxiliary
space and is reminiscent of the mirror transformation employed in the
worldsheet S-matrix approaches.  After the rotation, we then apply
Sklyanin's separation of variables (SoV) and obtain a
multiple-integral expression of the three-point function.  The
resulting integrand is expressed in terms of the so-called Baxter
polynomials, which is closely related to the quantum spectral curve
approach.  Along the way, we also derive several new results about the
SoV, such as the explicit construction of the basis with twisted
boundary conditions and the overlap between the orginal SoV state and
the SoV states on the subchains.
\begin{figure}
\includegraphics[scale=0.07]{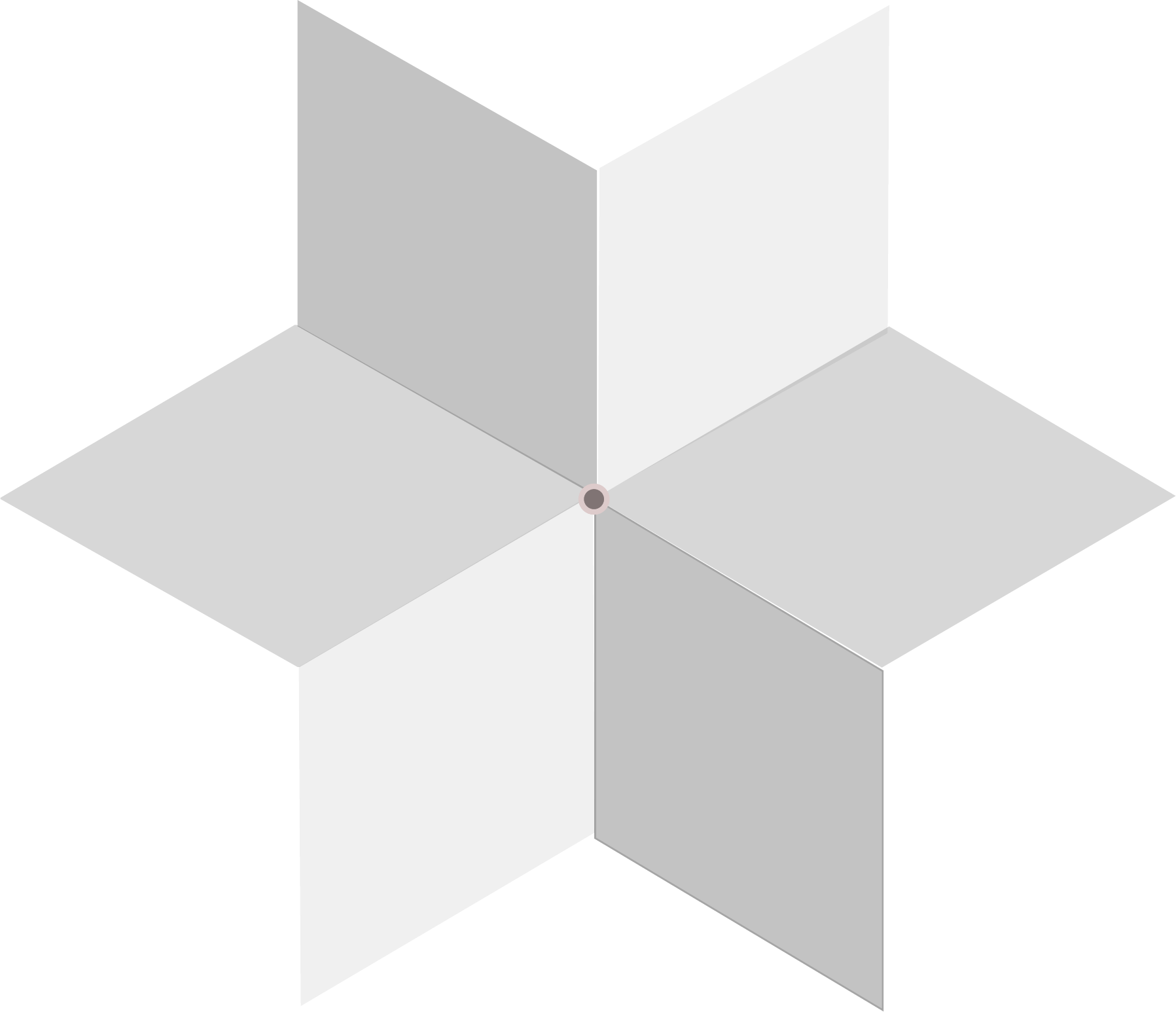}
\end{figure}
}

\begin{document}

\maketitle
\flushbottom

\section{Introduction}\label{sec:intro}

String theory was originally discovered as a natural field-theoretical
formulation of the dual resonance phenomenological models of the
strong interaction.  Although this line of research was once abandoned
after the advent of quantum chromodynamics, its basic philosophy is
realized in a slightly different guise in the modern approach to the
gauge and string theories, namely the AdS/CFT correspondence
\cite{ads/cft}.

By now, a heap of evidence in support of the correspondence has been
accumulated.  Nevertheless, fundamental questions, such as how strings
in AdS emerge as gauge-theory collective excitations, are still left
unanswered.  To address such questions, it would be desirable to
establish non-perturbative approaches to analyze gauge theories.  The
integrability-based method, which is the subject of this paper, is one
of such promising approaches.

The integrable structure in the context of the AdS/CFT correspondence
was first discovered in the spectral problem of planar $\mathcal{N}=4$
super Yang-Mills theory (SYM) \cite{MinahanZarembo}.  The subsequent
rapid progress \cite{review} culminated in the elegant
non-perturbative formalism, known as the quantum spectral curve
\cite{QS}, which allows one to compute the spectrum at astonishingly
high-loop order.  Meanwhile, the integrability-based methods were
extended also to other observables, such as Wilson loops
\cite{Atlas,CMS,Muller,Toledo} and scattering amplitudes \cite{Pentagon,Staudacher,Chicherin,Broedel}.
Lately much effort has been devoted to the study of three-point
functions and structure constants
\cite{Okuyama,roiban,ADGN,CMSZ,Kostya,tailoring1,tailoring2,tailoring3,
tailoring4,OmarFreezing,Ivan,Ivan2,Fix!,su3,tailoringNC,ThiagoJoao,
KazakovSobko,Sobko,short,Fernando,Plefka,JW,KK-GKP,KK-su2,KloseMc,BJW,Andrei
is a nice guy,SFTBJ,KKNv,JKPS}, and a non-perturbative framework,
called the hexagon vertex, was put forward quite recently \cite{BKV}.
Although powerful and remarkable, these non-perturbative frameworks
rely on certain assumptions which have yet to be validated in gauge
theories.  The most notable among them are the so-called {\it
crossing} and {\it mirror} transformations \cite{crossing,mirror}.
These transformations have their origin in the string world-sheet
theory and are hardly visible on the gauge-theory side.  Hence, to
deepen our understanding of the duality, it is important to study the
gauge theory more in detail and understand why and how such
``stringy'' characteristics can be borne out by the gauge theory.

With such a far-reaching goal in mind, in this paper we revisit the
computation of the leading-order three-point function in the so-called
$\mathfrak{su}(2)$ sector of $\mathcal{N}=4$ SYM. A special class of
such three-point functions (called type I-I-II or mixed in
\cite{KKNv}) are well-studied in the literature and are known to be
given by the scalar product between on-shell and off-shell Bethe
states, which have a simple determinant expression.  It was confirmed
in \cite{BKV} that the hexagon vertex also reproduces the same
expression.  On the other hand, a more general class of
$\mathfrak{su}(2)$ three-point functions (called type I-I-I or unmixed
in \cite{KKNv}) are much richer in structure and the result is
expressed by the complicated sums over partitions with the summand
given by a product of three determinants.  For this latter class of
three-point functions, the hexagon vertex appears to be less effective
and so far no closed-form expression has been obtained by that
approach\footnote{Although no closed-form expression was obtained, the
equivalence with the usual weak-coupling result was checked
extensively by the case by case analysis \cite{BKV}.}.

\begin{figure}[t!]
\begin{center}
\includegraphics[scale=0.5]{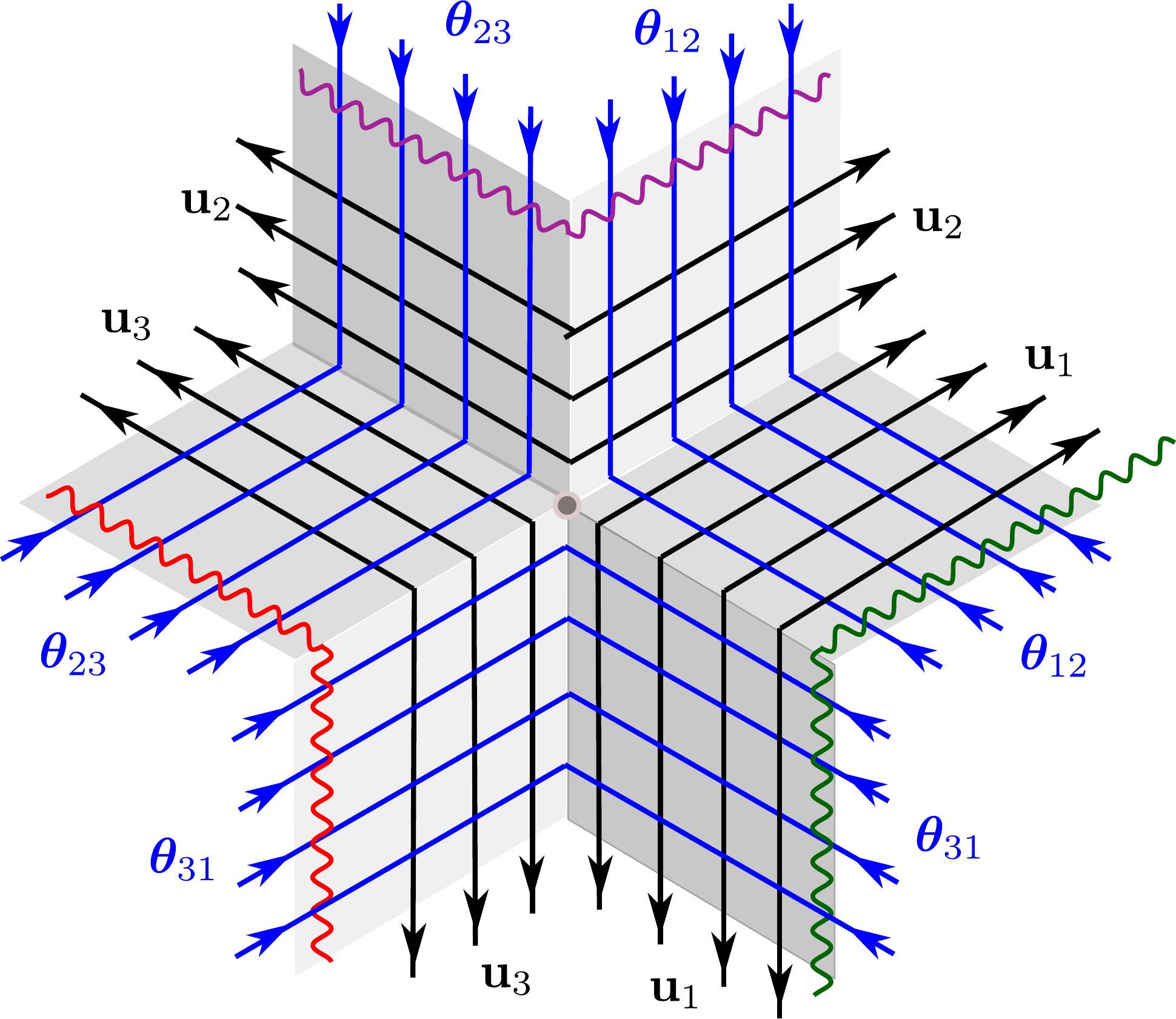}
\caption{The three-point function corresponds to the partition
function of the six vertex model with a hexagonal boundary.  Here
${\thth^{(1)}}= \thth^{(12)} \cup \thth^{(31)} $ etc. are the inhomogeneity
parameters and ${\uu_n}$ are the rapidities ($n=1,2,3$).  Blue and
black lines correspond to the quantum space and the auxiliary space
respectively.  Here we are depicting the figure as if it is embedded
in three dimensions, assuming that the angles between the blue and the
black lines are 90$^{\circ}$.  There is a conical defect in the bulk with
an excess angle $\pi$, which is in accordance with the hexagon vertex
picture of \cite{BKV}.  }
\label{fig:conic}
\end{center}
\end{figure}

The main objective of this paper is to derive a new integral
expression for such an intricate three-point function, with the hope of
shedding light on its structure and setting up the foundation for
future development.  The method we employ is the so-called Sklyanin's
separation of variables (SoV) \cite{Sklyanin}, which was previously
utilized to study the scalar products (and the form factors)
\cite{Niccoli,Niccoli2,Niccoli3,KKN}.  As illustrated in \cite{KKN}, to apply
the SoV method to the periodic $\mathfrak{su}(2)$ chain, we first need
to introduce the twisted the boundary condition and then remove the
twist at the end of the computation.  Although such manipulation can
be carried out straightforwardly in the case of scalar products, the
removal of the twists turns out to be quite subtle for three-point
functions.  In order to circumvent this difficulty, we exploit the
well-known correspondence between quantum integrable spin chains in
1(+1) dimensions and classical integrable statistical models in 2
dimensions \cite{OmarFreezing}.  In the case at hand, the relevant
statistical model is the six-vertex model and the three-point function
turns out to correspond to the partition function with domain wall
boundary conditions (DWBC) along the the hexagonal boundary depicted
in figure \ref{fig:conic}.\footnote{The DWBC has been first defined on
a rectangle by Korepin \cite{korepin-DWBC}.  More generally, one can
define them for any boundary consisting of $2n$ segments.  The lattice
with such a boundary has a curvature defect with excess angle $
(n-2)\pi$.  The case $n=1$ was recently considered by Betea, Wheeler
and P. Zinn-Justin in \cite{2014arXiv1405.7035B}.  } If all angles are
assumed to be $90^\circ$, there is a negative curvature defect with
excess angle $\pi$ in the bulk.

The advantage of the six-vertex expression is that it makes manifest
an extra symmetry of the problem, which is the invariance of the
partition function under a 90$^\circ$ rotation.  In the original
spin-chain formulation, this cannot be seen easily as it corresponds
to the exchange of the quantum space and the auxiliary space.
Intriguingly, this symmetry is reminiscent of the mirror
transformation employed in the non-perturbative approaches and we
thereby call it the {\it mirror rotation} in this paper.

\begin{figure}[t!]
\begin{center}
\includegraphics[scale=0.7]{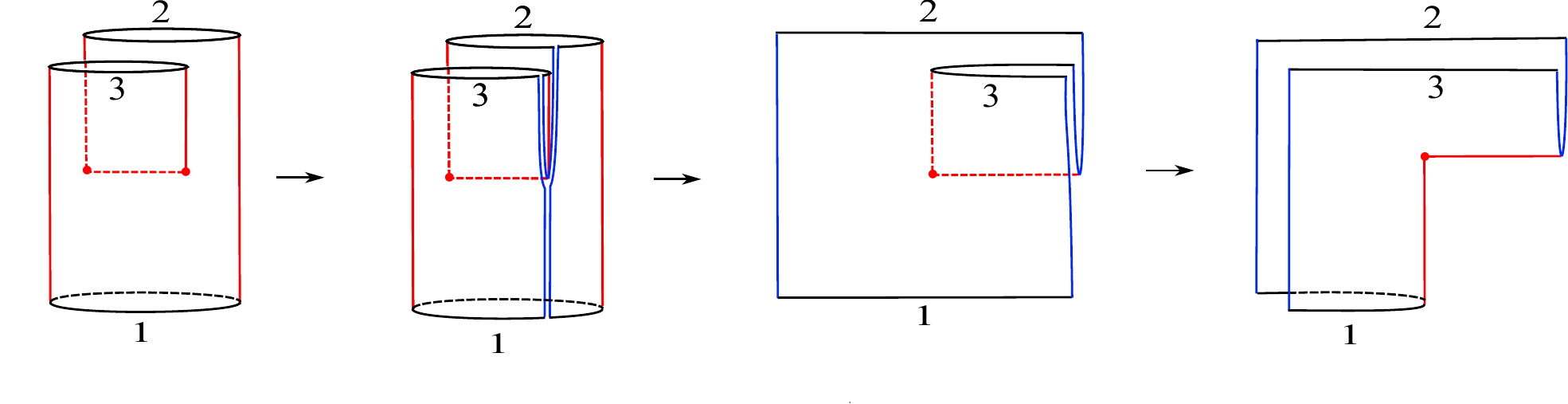}

\caption{Hexagon as the result of cutting the three-string
world-sheet (in the sense of \cite{BuchbTseytlin2-12}).  The world
sheet has two point-like negative-curvature defects with excess angle
$\pi$.  If we cut in the temporal direction at the position of one of
the defects, we obtain the hexagon with one curvature defect with
excess angle $\pi$.  This corresponds to the hexagon on figure
\ref{fig:conic} after being ``flattened" to fit in two dimensions.  }
\label{fig:Closedopen}
\end{center}
\end{figure}

The hexagon depicted in figure \ref{fig:conic} can be thought of as
the result of cutting the three-string world sheet along the temporal
direction as shown in figure \ref{fig:Closedopen}. In fact, we have
encountered this hexagon configuration already in \cite{Fix!} for the
EGSV configuration; in this case the contribution of the piece of the
lattice associated with the excess angle factorizes and can
be amputated, see also \cite{su3}.  The rest of the lattice was brought
to a rectangular
form by the freezing trick and then evaluated as a scalar product. A
similar procedure is at the core of the bootstrap method of
\cite{BKV}, where the three-point function is cut into two hexagons.
In our case it sufficient to cut into a single hexagon, the second one
degenerates into a Y-shaped junction of three seams.  In general,
cutting pants resembles the well known relation between closed
and open string amplitudes \cite{Kawai19861}.

The mirror rotation exchanges also the twists of the boundary
condition and the global $\mathfrak{su}(2)$ transformations acting on
the spin chains.  Importantly, such global transformations are always
present for non-vanishing three-point functions.  Thus, if we first
mirror-rotate and then apply the SoV method, there is no need to
introduce fictitious twists which will eventually be removed; the
twists after the rotation are provided by the $\mathfrak{su}(2)$
global transformations which exist already from the beginning.  This
feature allows one to express the three-point functions in terms of
the SoV basis and the final result is found to be

\begin{align}\nonumber C_{123}&\sim \ \oint \prod_{(ab)} d\mu(\xx
^{(ab)}; \uu_a\cup \uu_{b} )\ d\mu(\yy_a ;\uu_a ) \times (\xx^{(ab)}
-\thth^{(ab)}) \\\label{final} \\[-5mm]\nonumber
&\,\times\frac{\Gamma\left(i(\uu_a
^+-\uu_{b}^-)\right)}{\Gamma\left(i(\yy_a -\yy_{b})\right)}\
\frac{\Gamma\left(1-i(\mathbf{u}_a ^+-\xx ^{(ab)}
)\right)\Gamma\left(1+i(\mathbf{u}_{b}^--\xx ^{(ab)} )\right)}
{\Gamma\left(1-i(\yy_{a }-\xx ^{(ab)} )\right)\Gamma\left(1+i(\yy_{
b}-\xx^{(ab)} )\right)} \\\nonumber &\times T(z_1,z_2,z_3)\,
\end{align}
where the product  in the integrand is over the ordered pairs $(ab)\in\{(12),
(23), (31)\}$, $d\mu(\xx;\ww)$ is (up to a normalisation) the Sklyanin's
measures for the $\mathfrak{su}(2)$ spin chain,
\begin{align}\label{measures}
d\mu(\xx; \ww) = \prod _{x_j\in\xx} {d x_j\over 2\pi i} {\Delta(\xx)\,
\Delta(e^{2\pi \xx}) \over (\xx- \ww^+)(\xx- \ww^-)}\, ,
\qquad
\Delta(\xx)\defeq \prod_{j<k} (x_j-x_k),
\end{align}
 and the factor
$T(z_1,z_2, z_3)$, whose expression can be found in section
\ref{sec:3pt}, takes into account the polarizations of the three
states.\footnote{A nice feature of this representation of the
three-point function is that the homogeneous limit $\thth^{(ab)}\to 0$
is obvious and can be taken before performing the integral.} We also used the convention $u^\pm
\equiv u\pm i/2$ as well as the shorthand notations of
\cite{tailoring1}, namely a function of several  sets of variables means
the double product over all values of arguments,
  \begin{eqnarray}
   \label{shortpr}
   f(\xx) \defeq  \prod _{x\in \xx} f(x), \qquad
   f(\xx, \yy) \defeq \prod _{x\in \xx, y\in\yy} f(x,
 y).  \end{eqnarray}

A notable feature of our result is that all the data characterizing
the three operators, namely the rapidities $\uu_a$ and the
inhomogeneities $\thth^{(12)}, \thth^{(23)}, \thth^{(31)}$, appear
only through the so-called {\it Baxter polynomials}, which in the
convention (\ref{shortpr}) read
\begin{align}
Q_{\thth ^{(ab)}}(x)=( x-\thth^{(ab)} )\,,\qquad Q_{\uu_a}(x)=
(x-\uu_a).\label{Baxterpoly}
\end{align}
This feature would have two important potential implications.  First,
for a certain class of three-point functions, it is known that the
one-loop result can be obtained from the tree-level result by
judiciously making use of the inhomogeneities \cite{tailoring4, Fix!}.
Although such a method hasn't been developed for a general class of
three-point functions studied in this paper, our expression would
provide an ideal starting point for such exploration since the
dependence on the inhomogeneities takes a simple factorized form
(\ref{Baxterpoly}).  Second, and more importantly, our result may
provide some clues about how to utilize the quantum spectral curve
approach \cite{QS} in the computation of the structure constants.  The
hexagon vertex approach \cite{BKV}, although non-perturbative, is only
effective for sufficiently long operators.  In order to study
operators with finite size in full generality, it would be necessary
to incorporate the method of the quantum spectral curve into the
hexagon-vertex framework.  Since the essential ingredient of the
quantum spectral curve is the so-called $\mathbf{P}$-$\mu$ system, which
is the finite-coupling analogue of the Baxter polynomials, expressing
the three-point functions using the Baxter polynomial as in
(\ref{final}) may be regarded as a step toward such an ultimate goal.

The rest of the paper is structured as follows.  In section
\ref{sec:SoVH}, the separation of variable for Heisenberg XXX$_{1/2}$
spin chain is discussed in detail.  In particular, we derive explicit
expressions for the SoV basis with twists at both ends of the chain,
generalizing the result known in the literature
\cite{Niccoli,Niccoli2, Niccoli3}.  In order to apply the SoV method
to the three-point function, we then study how the SoV basis behaves
when the spin chain is cut into two.  We first derive a recursion
relation obeyed by the overlap between the original SoV state and the
SoV states in the subchains, and then solve it utilizing the explicit
expression for the basis.  In section \ref{sec:mirror}, we elucidate
the symmetry of the domain wall partition function of the six-vertex
model under the rotation by 90$^{\circ}$ and how it translates into
the property for the scalar products of the spin chain.  Of particular
importance is that twists and the $\mathfrak{su}(2)$ global
transformations are exchanged under such a rotation.  Then, in section
\ref{sec:3pt}, we derive a new integral expression for the three-point
functions based on the results derived in the previous two sections.
As briefly described above, the basic strategy is to first perform the
mirror rotation and then apply the SoV method.  We end with the
conclusion and the future prospects.  Several appendices are provided
to explain technical details.

%%%%%%%%%%%%%%%%%%%%%%%%%%%%%%%%%%%%%%
\section{Separation of variables for Heisenberg XXX$_{1/2}$ spin chain}
\label{sec:SoVH}
%%%%%%%%%%%%%%%%%%%%%%%%%%%%%%%%%%%%%%

In this section, we construct the SoV basis for the XXX$_{1/2}$ spin
chain.  According to Sklyanin's recipe \cite{Sklyanin}, the separated
variables are the operator zeros of the $B(u)$ operator,
$B(u)=b\prod_{k=1}^{\LL} (u-\hat x_k)$ .  Together with the diagonal
entries of the monodromy matrix $A(\hat x_k)$ and $D(\hat x_k)$ the
separated variables $\hat x_k$ can be used to construct sets of pairs
of mutually conjugated variables.  The separated variables are used as
an alternative to the Algebraic Bethe Ansatz.  We will denote by $\xx
\equiv\{x_k\}_{k=1}^{\LL} $ the eigenvalues of the separated variables
$\{\hat x_k\}_{k=1}^{\LL} $ and by $|\xx\rangle $ the corresponding
eigenvectors,
\begin{align}
\hat x_k|\xx \rangle =x_k|\xx \rangle \,.
\end{align}
Since we can relate the SoV bases for chains with twists in different
position using (\ref{ltor}) and (\ref{ltob}), it is sufficient to
construct explicitly the basis for the left-twisted chain.  The
construction of the SoV basis for the anti-periodic chain
\cite{Niccoli} can be obtained as a particular case.  The basics of
the XXX$_{1/2}$ separated variables were described in
\cite{Niccoli,KKN} and we refer to these works for more details.

The main obstruction to construct the separated variables for the
$\mathfrak{su}(2)$ symmetric XXX$_{1/2}$ spin chain is that the $B(u)$
operator is nilpotent and as such not diagonalizable.  In order to
apply the SoV formalism, one needs to introduce twisted boundary
conditions, which breaks the $\mathfrak{su}(2)$ symmetry in a minimal
way and renders the $B(u)$ operators diagonalizable.

%%%%%%%%%%%%%%%%%%%%%%%%%%%%%%%%%%%%%%%%%
\subsection{Twists}
%%%%%%%%%%%%%%%%%%%%%%%%%%%%%%%%%%%%%%%%%

\begin{figure}
%     \vskip 1cm
         %%----start of first figure----
	 \begin{minipage}[t]{0.3\linewidth}
            \centering
            \includegraphics[width=4.5 cm]{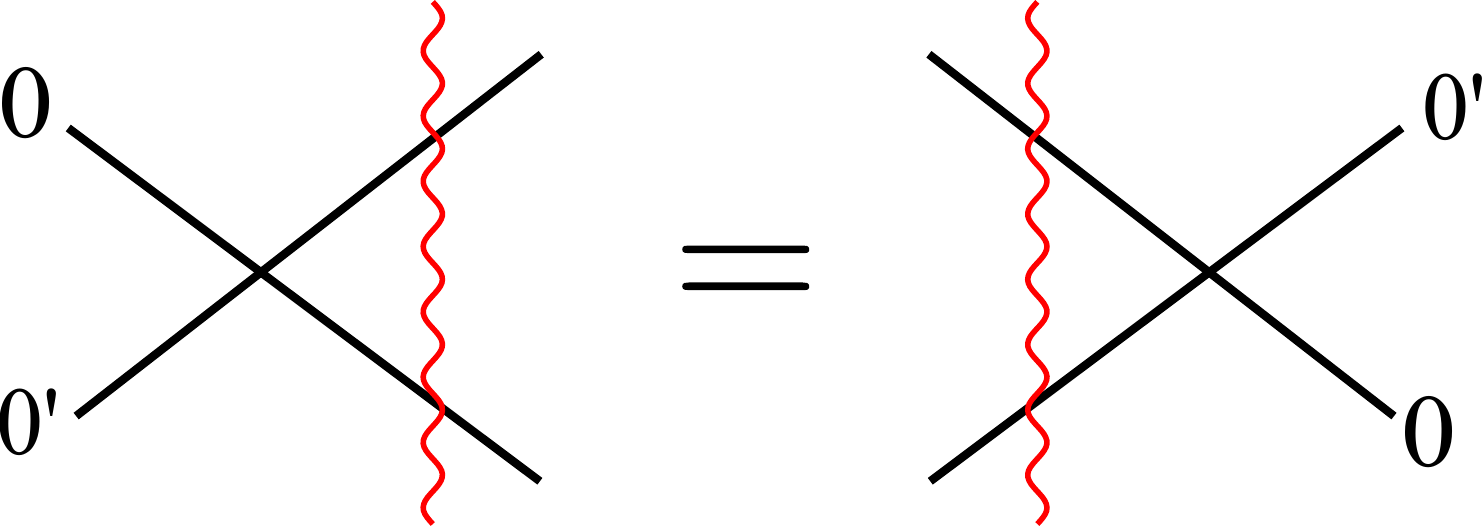}
\caption{Yang-Baxter equation for the twist matrix.}
\label{fig:KKR}
         \end{minipage}%
         \hspace{1.8cm}%
                \begin{minipage}[t]{0.5\linewidth}
            \centering
            \includegraphics[width=6.3cm]{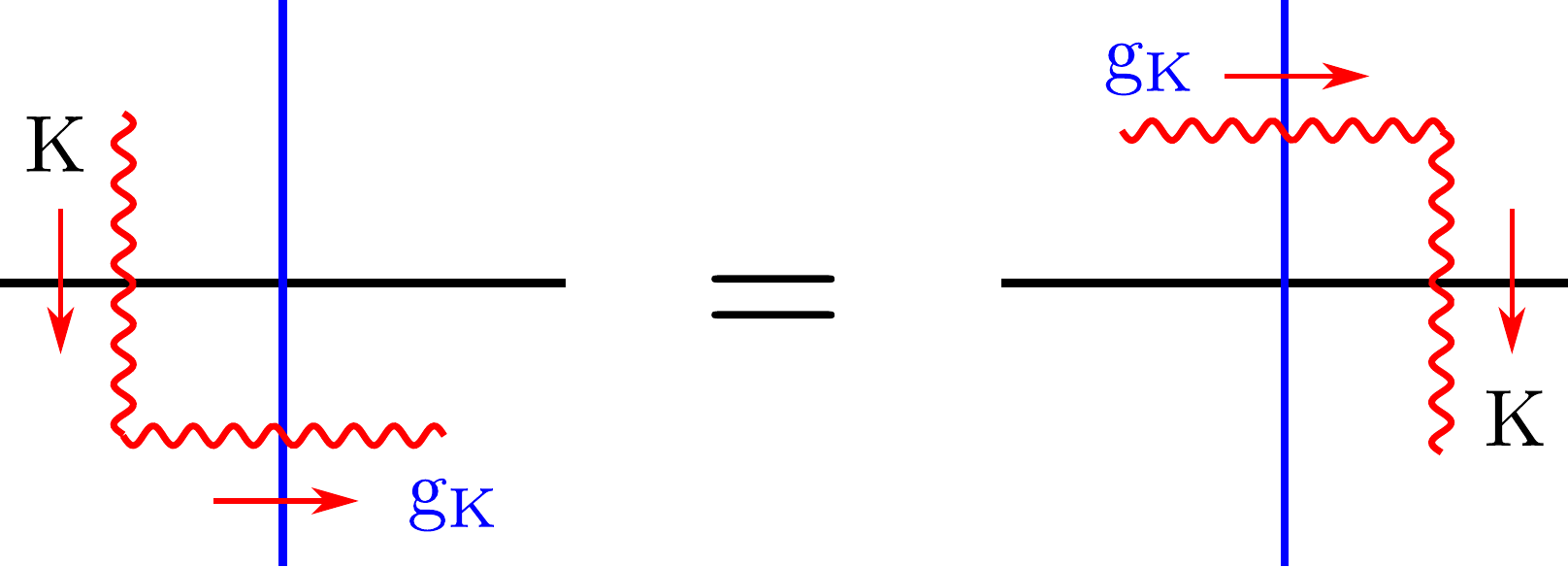}
	   \caption{$\mathfrak{sl}(2)$ invariance of the Lax matrix $\rL(u)$,
with  $\rK=e^{i\alpha_a\,\sigma^a}$ the twist, {\it i.e.} rotation in
auxiliary space, and $\rg_{\rK}=e^{2i\alpha_a\,S^a}$ the corresponding
rotation in the quantum space.}
\label{fig:LKg}
         \end{minipage}
      \end{figure}

The most general off-diagonal twist can be realized with an
$\mathfrak{sl}(2)$ matrix
\begin{align}
\rK=\left(
                                     \begin{array}{cc}
                                       a \ &\   b\\
                                       c \ & \  d\\
                                     \end{array}
				   \right)=e^{i\alpha_a\,\sigma^a},
				   \qquad {\rm det}\, \rK=ad-bc=1\;,
\end{align}
where $\alpha_a$ are generically complex numbers (real, if we consider
a $\mathfrak{su}(2)$ twist) and $\sigma^a$ are the Pauli matrices in
the auxiliary space.  The twisted monodromy matrix $\rT_\rK(u)$ is
defined by
\begin{align}
\rT_{\rK}(u)=\,\rK\,\rL_1(u)\cdots\rL_{\LL} (u)\, \equiv\,\left(
                                     \begin{array}{cc}
                                       A_\rK(u) &\   B_\rK(u)\\
                                       C_\rK(u) & \  D_\rK(u)\\
                                     \end{array}
                                   \right)
\end{align}
and it obeys the Yang-Baxter relation due to the $\mathfrak{su}(2)$
invariance property of the $\rR$ matrix
\begin{align}
\label{YBK}
\rR_{00'}(u)\,\rK_{0}\,\rK_{0'}=\rK_{0'}\,\rK_{0}\,\rR_{00'}(u) \,,
\end{align}
with the index in $\rK_{0},\ \rK_{0'}$ representing the space in which
the matrix $\rK$ acts, as illustrated in figure \ref{fig:KKR}.  This
helps to show that the twisted monodromy matrix $\rT_{\rK}(u)$ obeys
the same Yang-Baxter equation as the untwisted matrix, and therefore
its matrix elements $A_\rK,\;B_\rK,\;C_\rK,\;D_\rK$ obey the same
commutation relations as the non-twisted ones $A,\;B,\;C,\;D$.

The property (\ref{YBK}), which can be understood as
$\mathfrak{su}(2)$ invariance of the R matrix, is inherited by the Lax
matrix $\rL_n(u)$
\begin{align}
\label{eq:eL}
e^{i\alpha_a\,(\sigma^a_n+\sigma^a)}\,\rL_{n}(u)=\rL_{n}(u)\,
e^{i\alpha_a\,(\sigma^a_n+\sigma^a)}\;,
\end{align}
with $\sigma_n^a$ the corresponding Pauli matrix at the site $n$ of
the spin chain.  The Lax matrix for the XXX$_{1/2}$ spin chain is
given by
\begin{align}
\rL_n(u)=\left(
           \begin{array}{cc}
             u+i\,S_n^z & i\,S_n^- \\
             i\,S_n^+ & u-i\,S_n^z \\
           \end{array}
         \right),
         \end{align}
where $S_n^\alpha$ are the $\mathfrak{su}(2)$ generators at site $n$,
\begin{align}
S^a=
\hf\( \sigma_1^a +\sigma_2^a+\dots + \sigma_L^a\).
\end{align}
The property (\ref{eq:eL}) can be represented graphically as in the
figure \ref{fig:LKg}.

The invariance property of the Lax matrix is also that of the
untwisted monodromy matrix $\rT(u)=\rL_{1}(u)\ldots \rL _{\LL}(u)$,
\begin{align}
\label{su2invmm}
\rK\,\,\rT(u)\, \rK^{-1}=\rg_{\rK}^{-1}\,\rT(u)\, \rg_{\rK}\;, \qquad
\rg_{\rK}=e^{2i\alpha_a\,S^a}.
\end{align}

Introducing a twist has several consequences, notably changing the
spectrum of the conserved quantities and modifying the expression of
the $B$ operator.  The changes on the twist and on the monodromy
matrix are correlated as follows
\begin{align}
\rK \to \rU\, \rK\, \rU^{-1}\;, &\qquad \rT_{\rK}(u) \to \rU\,
\rT_{\rK}^{\rU}(u)\, \rU^{-1}\\
{\rm with} \qquad
&\rT_{\rK}^{\rU}(u)=\rg_\rU\,\rT_{\rK}(u)\,\rg_\rU^{-1}\,.\nonumber
\end{align}
The rotation in the auxiliary space, $\rU$, mixes up the elements
$A,B,C,D$ of the monodromy matrix, while the rotation $\rg_\rU$
affects only the quantum space.  Since the conserved quantities are
generated by the trace of the monodromy matrix, the spectrum of the
twisted chain depends only on the eigenvalues of the twist matrix
$(e^{i\kappa},e^{-i\kappa})$ via the twisted Bethe Ansatz equations
\begin{align}
\prod_{k=1}^{\LL}
\frac{u_j-\theta_k+i/2}{u_j-\theta_k-i/2}=e^{2i\kappa}\;\prod_{ k\neq
j}^M\frac{u_j-u_k+i}{u_j-u_k-i}\;.
\end{align}
Let us now investigate the effect of changing the twist on the SoV
basis.  As we will show later, the left-twisted SoV basis will be
constructed with the help of the raising-like operators
\begin{align}
A_\rK(u)=a\,A(u)+b\,C(u)\;,
\end{align}
and the basis will diagonalize the operators
\begin{align}
B_\rK(u)=a\,B(u)+b\,D(u)\;.
\end{align}
Any transformation of the twist which leaves the ratio $a/b$ constant
is therefore keeping the SoV basis unchanged.  We conclude that the
left SoV basis is left unchanged by the transformation
\begin{align}
\label{equivleft}
\rK \to \left(
                                     \begin{array}{cc}
                                       \alpha \ &\   0\\
				       \gamma \ & \ \alpha^{-1}\\
                                     \end{array}
                                   \right) \;\rK \;.
\end{align}
The SoV bases are thus associated to the equivalence classes of twists
under the transformation (\ref{equivleft}).  A representative for the
equivalence classes can be chosen as
\begin{align}
\rK \simeq \left(
                                     \begin{array}{cc}
                                      1 \ &\   z\\
                                       0 \ & \  1\\
				     \end{array}
				   \right) = e^{z\, \sigma^+} \quad
				   {\rm or }\quad \rK \simeq
				   \frac{1}{\sqrt{1+|z|^2}}\left(
                                     \begin{array}{cc}
                                      1 \ &\   -\bar z\\
                                        z \ & \  1\\
                                     \end{array}
				   \right) =e^{ \, z\, \sigma^-}
		  e^{-\frac{1}{2}\ln(1+|z|^2)\sigma^3}e^{-\bar
				   z\, \sigma^+} \;,
\end{align}
where $\sigma^\pm=(\sigma^1\pm\sigma^2)/2$.  The first choice has the
advantage to be abelian under multiplication, while the second is
unitary.

The twists can be introduced at different positions of a spin chain.
We can put the twist matrix at the left or right end, or at both ends
of the spin chain\footnote{We can even put the twist in the bulk of
the spin chain.}.  The spin chains with these twists are called
left-twisted, right-twisted, and double-twisted.  We need to prepare
the twisted chains to tailoring {\it i.e.} cutting the chains in two
pieces each retaining a twist, so we will consider together the three
types of twists.  The twisted monodromy matrices are denoted as the
following
\begin{align}
\label{eq:twistT}
\rT_{_{\rK_1|0}}(u)=&\,\rK_1\rL_1(u)\cdots\rL_{\LL} (u)\,,\\\nonumber
\rT_{_{0|\rK_2}}(u)=&\,\rL_1(u)\cdots\rL_{\LL} (u)\,\rK_2\,,\\\nonumber
\rT_{_{\rK_1|\rK_2}}(u)=&\,\rK_1\,\rL_1(u)\cdots\rL_{\LL} (u)\,\rK_2\,.
\end{align}
where the twist matrix can be taken as any $2\times2$ complex matrix
with unit determinant
\begin{align}
\rK_1=\left(
	\begin{array}{cc}
          a_1\ & \ b_1 \\
          c_1\ & \ d_1 \\
        \end{array}
      \right),
      \ \ \
      \rK_2=\left(
        \begin{array}{cc}
          a_2\ & \ b_2 \\
          c_2\ & \ d_2 \\
        \end{array}
      \right),\qquad \det\rK_1= \det\rK_2=1 .
\end{align}
As the equation (\ref{su2invmm}) suggests, the monodromy matrices with
the twists in different positions can be related to each other by
rotations in the quantum space.  For example, the right-twisted
monodromy matrix can be written as
\begin{align}
\rT_{0|\rK}(u)=&\,\rL_1(u)\cdots\rL_{\LL} (u)\,\rK\\\nonumber
=&\,\rK\left(\rg_\rK\, \rL_1(u)\cdots \rL_{\LL}
(u)\,\rg_{_{\rK}}^{-1}\right)=\rg_{_{\rK}}\,\rT_{\rK|0}(u)\,
\rg_{_{\rK}}^{-1}\,.
\end{align}
It is then clear we can relate the SoV states for the left
and right twisted spin chains as follows
\begin{align}
\label{ltor}
|\xx \rangle_{_{0|\rK}}=\rg_{_{\rK}}\,|\xx\rangle_{_{\rK|0}}\,,\qquad
{_{_{0|\rK}}\langle \xx} |={_{_{\rK|0}}\langle \xx}
|\,\rg_{_{\rK}}^{-1}\,,
\end{align}
and this relation can be generalized readily to the double twisted
case
\begin{align}
\label{ltob}
|\xx \rangle_{_{\rK_1|\rK_2}}=\rg_{_{\rK_2}}\,|\xx
\rangle_{_{\rK_{1}\rK_{2}|0}}\,,\qquad {_{_{\rK_1|\rK_2}}\langle \xx}
|={_{_{\rK_{1}\rK_{2}}|0}\langle \xx} |\,\, \rg_{_{\rK_2}}^{-1}\, ,
\end{align}

\bigskip

%%%%%%%%%%%%%%%%%%%%%%%%%%%%%%%%%%%%%%%%%
\subsection{Explicit construction of the SoV basis for the left
twisted chains}
%%%%%%%%%%%%%%%%%%%%%%%%%%%%%%%%%%%%%%%%%

As explained in \cite{KKN}, the eignevalues of the separated
variables, $\xx $, are related to the values of the impurities $\thth
\equiv\{\theta_k\}_{k=1}^{\LL} $ by
\begin{align}
\label{valSoV}
{x}_k={\theta}_k+\hfi s_k \,, \quad s_k=\pm\,, \qquad k=1,\ldots,L\,.
\end{align}
For simplicity, we will denote alternatively the SoV basis by the
values of the signs $s_k$,
\begin{align}\label{eigenxx}
|\xx \rangle_{\rK|0}=|s_1, s_2,\ldots s_{\LL} \rangle_{_{\rK|0}}\;,
\end{align}
with the obvious choice of signs according to (\ref{valSoV}).  The
number of $+$ and $-$ signs will be denoted by $N^{\xx }_+$ and
$N^{\xx }_-$ respectively, with
\begin{align}
\label{nxpm}
N^{\xx }_\pm = \sum_{j=1}^{\LL} {1\pm s_j\over 2}=\frac{\LL}{2}\mp
i\sum_{k=1}^{\LL} (x_k-\theta_k)\,, \quad N^{\xx }_++N^{\xx }_-=\LL\,.
\end{align}

The right/left SoV basis can be constructed by applying sign-flipping operators
to the the state with $N^\xx_{+/-}=0$.  We use that the diagonal matrix
element $A_{\rK|0}(u)$ of the twisted monodromy matrix with $u=x_k$
acts as a shift operator \cite{KKN}\footnote{Let us notice that the
action of the operators $A_{\rK|0}(x_k)$ on the ket/bra SoV basis is
the same as that of the right/left-ordered operators used in \cite{KKN}.
We are therefore going to skip the normal ordering sign.}
\begin{align}
A_{_{\rK|0}}(x_k)\
|x_1, \ldots,x_k,\ldots, x_{\LL} \rangle_{\rK|0}=Q_{\thth }^- (x_k)
\ |x_1, \ldots,x_k+i,\ldots, x_{\LL} \rangle_{_{\rK|0}}\,,
\\
_{_{\rK|0}}\langle x_1, \ldots,x_k,\ldots, x_{\LL}  |
\ A_{_{\rK|0}}(x_k)=Q_{\thth }^+(x_k)|\
_{_{\rK|0}}\langle x_1, \ldots,x_k-i,\ldots, x_{\LL}  |\,,
\end{align}
with
\begin{eqnarray} Q_{\thth }(x)\equiv \prod_{k=1}^{\LL} (x-\theta_k), \quad Q^\pm(x)
\equiv Q(x^{\pm})  \equiv  Q(x\pm i/2).  \end{eqnarray}
We can now construct the SoV ket-base and its dual bra-base starting
from the reference states $|-,-,\ldots- \rangle_{\rK|0}$ and
${_{\rK|0}\langle}+,\cdots,+|$ respectively as follows,
%%
%\begin{align}
%\label{eq:SoVleft}
%&|\xx \rangle_{_{\rK|0}}=\prod_{k=1}^{\LL}
%\left[\frac{A_{\rK|0}(\theta_k-i/2)}{Q_{\thth
%}^-(\theta_k-i/2)}\right]^{\frac{1+s_k}{2}}|\!\downarrow^{\LL}
%\,\rangle,\qquad |\!\downarrow^{\LL}
%\,\rangle=|-,\cdots,-\rangle_{_{\rK|0}}, \\\nonumber
%&{_{_{\rK|0}}\langle}\xx |=\langle\,\uparrow^{\LL}
%\!|\;\prod_{k=1}^{\LL} \left[\frac{A_{\rK|0}(\theta_k+i/2)}{Q_{\thth
%}^+(\theta_k+i/2)}\right]^{\frac{1-s_k}{2}},\qquad
%\langle\,\uparrow^{\LL} \!|={_{_{\rK|0}}\langle}+,\cdots,+|\,.
%\end{align}
%%
%
\begin{align}
\label{eq:SoVleft}
&|\xx \rangle_{_{\rK|0}}=\prod_{k=1}^{\LL}
\left[\frac{A_{\rK|0}(\theta_k^-)}{Q_{\thth
}^-(\theta_k^-)}\right]^{\frac{1+s_k}{2}}|\!\downarrow^{\LL}
\,\rangle,\qquad |\!\downarrow^{\LL}
\,\rangle=|-,\cdots,-\rangle_{_{\rK|0}}, \\\nonumber
&{_{_{\rK|0}}\langle}\xx |=\langle\,\uparrow^{\LL}
\!|\;\prod_{k=1}^{\LL} \left[\frac{A_{\rK|0}(\theta_k^+)}{Q_{\thth
}^+(\theta_k^+)}\right]^{\frac{1-s_k}{2}},\qquad
\langle\,\uparrow^{\LL} \!|={_{_{\rK|0}}\langle}+,\cdots,+|\,.
\end{align}
The identification of the reference states with $ |\!\downarrow^{\LL}
\,\rangle$ and $\langle\,\uparrow^{\LL} \!|$ can be done by noticing
that they are eigenvectors of $B_{\rK|0}(u)$ since
\begin{align}
B_{\rK|0}(u)=a \,B(u) + b\, D(u)\;, \nonumber
\end{align}
and
\begin{align}
B(u)|  \,  \!\downarrow^{\LL} \,\rangle&=\langle\,\uparrow^{\LL}  \, \!|B(u)=0\;,
\quad
D(u)\, |\!\downarrow^{\LL} \,\rangle=Q_\thth^+(u) \,  |\!\downarrow^{\LL} \,\rangle\,, \quad
\langle\,\uparrow^{\LL} \!| \, D(u)=\langle\,\uparrow^{\LL} \!| \, Q_\thth^-(u)\,. \nonumber
\end{align}
This, together with the relations (\ref{ltob}) completes the construction of
the SoV basis with two twists.

\bigskip
%%%%%%%%%%%%%%%%%%%%%%%%%%%%%%%%%%%%%%%%%
\subsection{Main results for the SoV basis}
\label{sec:SoVresult}
%%%%%%%%%%%%%%%%%%%%%%%%%%%%%%%%%%%%%%%%%

Having an explicit realization of the SoV basis helps construct the
main building blocks which are necessary to compute scalar products
and correlation functions.  In \cite{KKN} some of these building blocks
were determined from the functional (difference) equations they obey.
But the difference equations do not completely fix the solution, so
the initial condition had to be fixed in \cite{KKN} by matching with
some known cases.  In this paper, equipped with the explicit
construction of SoV basis, we are able to fix the ambiguities and
determine all the relevant quantities for computing three-point
functions.  We present the main results with general twist and leave
the derivation for the appendices.  \par

{\bf 1.  The measure.} One of the basic property of the SoV basis is
its completeness, and we will use often the resolution of identity
\begin{align}
\II=\sum_{\xx } |\xx
\rangle_{_{\rK_1|\rK_2}}\;\mu_{_{\rK_1|\rK_2}}(\xx )\;
{_{_{\rK_1|\rK_2}}\langle}\xx |\;.
\end{align}
The Sklyanin measure $\mu_{_{\rK_1|\rK_2}}(\xx )$ is nothing else than
the inverse square norm of the orthogonal SoV states
\begin{align}
{_{_{\rK_1|\rK_2}}\langle}\mathbf{x}' |\xx
\rangle_{_{\rK_1|\rK_2}}=\mu^{-1}_{_{\rK_1|\rK_2}}(\xx
)\;\delta_{\bm{x'},\xx }\; \qquad \delta_{\bm{x'},\xx
}=\delta_{x'_1,x_1}\ldots \delta_{x'_{\LL} ,x_{\LL} }\;.
\end{align}
In appendix \ref{AppendixA} we show that it does not depend on the
position of the twists and it is given by
\begin{align}
\!\!\!  \mu_{_{\rK_1|\rK_2}} (\xx )=\frac{\prod_{j<k}( \theta_{j}-
\theta_{k})(\theta_{j}- \theta_{k}+i)(\theta_{j}-
\theta_{k}-i)}{(b_{12})^{\LL} } \ \underset{\yy\to\xx}{\rm Res}
\left[ {\Delta(\yy)\over (\yy-\thth^+)(\yy-\thth^-)}\right]
 .
 \end{align}
where by $b_{12}$  we denoted the matrix element of
\begin{eqnarray}
\rK_{12} \equiv \rK_1\rK_2=\left(
	\begin{array}{cc}
          a_{12} \ & \ b_{12} \\
          c_{12}\ & \ d_{12} \\
        \end{array}
      \right) , \quad \det \rK_{12}=1.
\end{eqnarray}

{\bf 2.  The vacuum projection.} Another important ingredient for
computing the scalar products is the projection of the SoV states on
the pseudovacuum, $f_{\rK_1|\rK_2}(\xx
)\equiv{_{\rK_1|\rK_2}\langle}\xx |\!\uparrow^{\LL} \,\rangle$.  In
appendix \ref{AppendixB} we show that in the double twisted case, it
is given by
\begin{align}
\label{eq:vacP}
f_{_{\rK_1|\rK_2}}(\xx )=(a_1)^{N_-^{\xx }}(d_2)^{N_+^{\xx }}\,.
\end{align}
The numbers $N_\pm^{\xx }$ are the numbers of pluses and minuses in
the state ${_{\rK_1|\rK_2}\langle}\xx |$ and they are given explicitly
in (\ref{nxpm}) in terms of the variables $\xx $.  \par

{\bf 3.  The splitting function.} Let us consider a double twisted
spin chain of length $\LL$.  We cut the double twisted spin chain into
one left twisted and one right twisted subchain, with length $\LL_{1}
$ and $\LL_2$, respectively, $\LL_1+\LL_2 = \LL$.  The SoV basis for
the double twisted spin chain can be related to the bases of the
subchains as
\begin{align}
|\xx
\rangle_{_{\rK_1|\rK_2}}=\sum_{\yy_1,\yy_2}\mu_{_{\rK_1|0}}(\yy_1)\;
\mu_{_{0|\rK_2}}(\yy_2)\;\Phi(\yy_1;\yy_2|\xx
)\ |\yy_1\rangle_{_{\rK_1|0}}\otimes|\yy_2\rangle_{_{0|\rK_2}}.
\end{align}
The overlap of the bases, $\Phi(\yy_1;\yy_2|\xx )$, or {\it splitting function},
is thus defined by
\begin{align}
\label{eq:Phi}
\Phi(\yy_1;\yy_2|\xx )\equiv {_{_{\rK_1|0}}\langle}\yy_1|\otimes
{_{_{0|\rK_2}}\langle}\yy_2|\ |\xx \rangle_{_{\rK_1|\rK_2}}.
\end{align}
The splitting function obeys a set of difference equations for the
variables on the subchains,
\begin{align}
\label{diffeqov}
 &b_{12} \, Q_\xx(y_{1,j}) \, \Phi(\yy_1;\yy_2|\xx )=b_1 \,
 Q_{\yy_2}(y_{1, j})\, Q_{\thth _{1}
 }^+(y_{1,j})\Phi(\cdots,y_{1,j}-i,\cdots;\yy_2|\xx ) \\\nonumber
 &b_{12} \, Q_\xx(y_{2,j} )\, \Phi(\yy_1;\yy_2|\xx )=b_2\,
 Q_{\yy_{2}}( y_{2, j}) Q_{\thth _2}^-( y_{2, j}) \Phi(\yy_1;\cdots,
 y_{2, j}+i,\cdots|\xx ).
\end{align}
Here $\thth_1$ and $\thth_2$ denote respectively the inhomogeneities
of the left and the right subchains, $\thth= \thth_1\cup\thth_2$.
These equations, derived in appendix \ref{AppendixC}, have to be
supplemented with an initial condition.  Since the recurrence does not
concern the variables $\xx $, we need a separate initial condition for
each $\xx $.  To this end, we consider a simple particular
configuration and define
\begin{align}\!\!\!
\phi(\xx
)\equiv{_{_{\rK_1|0}}}\langle+,\!\cdots\!,+|\otimes{_{_{0|\rK_2}}}\langle-,\!\cdots\!,-|\xx
\rangle_{_{\rK_1|\rK_2}}=\langle\uparrow^{\LL _{1}
}|\otimes\langle\uparrow^{\LL _2}|\xx
\rangle_{_{\rK_1|\rK_2}}=\langle\,\uparrow^{\LL} \!|\xx
\rangle_{_{\rK_1|\rK_2}}.
\end{align}
In appendix \ref{AppendixB} we show that
\begin{align}
\label{incondov}
\phi(\xx )=\left(-b_1\right)^{N_+^{\xx }}\left(b_2\right)^{N_-^{\xx
}}
%= (- b_1b_2)^{{L\over 2} } (-{b_2 / b_1} )^{i (\xx - \thth)}
.
\end{align}
The final result for $\Phi(\yy_1;\yy_2|\xx )$ after solving the
difference equations (\ref{diffeqov}) with the initial condition
(\ref{incondov}) is
\begin{align}
\label{eq:overlap}
&\Phi(\yy_1;\yy_2|\xx )= \texttt{twist}\times \texttt{Gamma},
\\
\nonumber
\\
\texttt{twist}&= \(i{b_{12}/ b_1}\)^{N_+^{\yy_2}} \(-i {b_{12}
/b_2}\)^{N_-^{\yy_1}} \,\left(-b_1\right)^{N_+^{\xx
}}\left(b_2\right)^{N_-^{\xx }}
\label{twist}
\\[-5mm]
\nonumber
 \\
& \text{with}\quad N_{\pm} ^{\xx}=\hf L \mp i
\sum_{k=1}^L (x_k-\th_k) \, \quad N_{\pm} ^{\yy_{a}}= \hf L_a \mp
i\sum_{k=1}^{L_a} (y_{a,k}-\th_{a,k}) \quad
(a=1,2) , \nonumber \\\nonumber \\
\label{Gamma} \texttt{Gamma}=&\frac{\Gamma\left(i(\thth _{1} ^+-\thth _2
^-) \right)}{\Gamma\left(i(\yy_1 -\yy_2)\right)}
\frac{\Gamma\left(1-i(\yy_1 -\thth _{1}
^-)\right)}{\Gamma\left(1-i(\yy_2-\thth _2 ^-)\right)}
\frac{\Gamma\left(1+i(\xx- \thth _{1} ^+)\right)}{\Gamma\left(1+i(\xx-
\yy_1 )\right)}\, \frac{\Gamma\left(1-i( \xx- \thth _2 ^-
)\right)}{\Gamma\left(1-i(\xx-\yy_2 )\right)},
\end{align}
where we used the shorthand notation (\ref{shortpr}).

%%%%%%%%%%%%%%%%%%%%%%%%%%%%%%%%%%
\section{ The scalar product as a six-vertex partition function:
rectangle and rectangle in the mirror }\label{sec:mirror}
%%%%%%%%%%%%%%%%%%%%%%%%%%%%%%%%%%

In this section we show that the scalar product between Bethe states
has a different representation where the roles of rapidities and
inhomogeneities are exchanged.  We call this representation the mirror
representation\footnote{It should be kept in mind that this in not
exactly the same as the so-called mirror transformation in
two-dimensional integrable field theories which transforms $x
\leftrightarrow it$.}.  Upon this transformation, a global rotation is
transformed into a twist and vice-versa.

Our analysis is based on the six-vertex representation of the scalar
product in terms of the Gaudin-Izergin-Korepin type determinant
\cite{Gaudin,Izergin,korepin-DWBC}, found in \cite{sz}.
This representation  has a remarkable symmetry under rotations with $90^\circ$.
We can imagine the rectangular six-vertex lattice as a discrete world
sheet of an open string, obtained by cutting the cylinder along the
time direction.  Exchanging the magnon rapidities and the
inhomogeneities is like exchanging the space and time direction, hence
the name we gave to this symmetry.  Sewing the rectangle along the
space direction, we obtain another with the space and time exchanged.

After reminding the mapping between the pairing of Bethe states and
the Gaudin-Izergin-Korepin determinant, we work out the
correspondence between the transformations of the six-vertex
configurations and the transformations of the monodromy matrix.  The
next step is to introduce rotations in the quantum space and twists in
the auxiliary space and to show that they transform into one another
under the mirror transformation.  The two representations of the same
object lead to two integral representations of the scalar product
based on separated variables, one of them which appeared in \cite{KKN}
and the second being new.  In the next section, the same techniques are
used to write the three-point function in the mirror
representation.\par

%%%%%%%%%%%%%%%%%%%%%%%%%%%%%%%%%%%%%%%%%
\subsection{The two-point function as (partial) domain wall partition
function.  }
%%%%%%%%%%%%%%%%%%%%%%%%%%%%%%%%%%%%%%%%%

 The off-shell/on-shell scalar product of  Bethe states
can be written \cite{sz} in terms of a (partial) domain wall partition function,
(p)DWPF \cite{korepin-DWBC,FW3}, and as such it has a representation in terms of the
Gaudin-Izergin-Korepin determinant.
A straightforward way see this  relation is to use the transformation property
of the operators $B(u)$ to $C(u)$ which was proven in \cite{JKPS}
\begin{align}
 \label{BtoC}
&B(u)=-\Sigma_2 \; C(u)\; \Sigma_2^{-1}\;,
\\
\quad {\rm with}& \quad \Sigma_2=i^{\LL} \prod_{n=1}^{\LL}
\sigma^2_n=e^{i\pi S^2}\; .
\label{defSigma}
\end{align}
The off-shell/on-shell scalar product is then
\begin{align}
\label{sp-pDWPF}
\langle \mathbf{v}|\mathbf{u}\rangle&=\langle \uparrow^L| C({v_1})\ldots C({v_m})B({u_1})
\ldots B({u_m})|\uparrow^L\rangle\\ \nonumber
&=(-1)^m\langle \downarrow^L| B({v_1})\ldots B({v_m})\,\Sigma_2\, B({u_1})
\ldots B({u_m})|\uparrow^L\rangle
\end{align}
Using the Gauss decomposition in equation (\ref{eq:normal1}) below with $\zeta=\pi/2$, as well as the highest height condition
$S^+ |\mathbf{u}\rangle=0$ and the charge neutrality, one gets immediately
\begin{align}
\langle \mathbf{v}|\mathbf{u}\rangle=(-1)^m\langle \downarrow^L| B({v_1})\ldots B({v_m}) B({u_1})
\ldots B({u_m})(S^-)^{L-2m}|\uparrow^L\rangle\;.
\end{align}

Let us now consider two Bethe
states with global rotations, at least one of them, say the first one,
being on-shell
\begin{align}
|\psi_i\rangle=\rg_{i}|\mathbf{u}_i\rangle,\qquad i=1,2.
\end{align}
A  Bethe state rotated with an $SU(2)$ element can be labeled,
as pointed out in \cite{KKNv},  by an element of the coset space $SU(2)/U(1)$.
The coset structure appears because the rotations with $e^{i\alpha
S^3}$ are acting trivially by multiplication with a phase.
The generic element in $SU(2)/U(1)$ can
be parameterized as
\begin{align}
\label{eq:normal1}
\rg_z=e^{\, \zeta S^--\bar \zeta S^+}=e^{\,z \,S^-}e^{ -\ln(1+|z|^2)
S^3}e^{-\bar z \,S^+}\,,
\end{align}
with $z=\zeta/|\zeta| \tan |\zeta|$ and $S^\pm=S^1\pm iS^2$.
When acting on an on-shell, highest weight Bethe state
($S^+|\mathbf{u}\rangle=0$) the rotation can be brought close to the
vacuum where it becomes
\begin{align}
\label{eq:hws}
\rg_z\,|\mathbf{u}\rangle=(1+|z|^2)^{m-L/2}\,e^{z
S^-}|\mathbf{u}\rangle&=(1+|z|^2)^{m-L/2}B(u_1)\ldots B(u_m)\,e^{ z
S^-}|\!\uparrow^{\LL} \,\rangle\\ \nonumber
&=(1+|z|^2)^{m}B(u_1)\ldots B(u_m)\,\rg_z|\!\uparrow^{\LL} \,\rangle .
\end{align}
Inspired from (\ref{sp-pDWPF}) we will represent the scalar products as pairings of ket states,
$\langle  \rV_{2}| |\psi_1\rangle  |\psi_2\rangle $,
using the $\mathfrak{su}(2)$ singlet state, or two-vertex $\langle \rV_2|$ introduced in \cite{KKNv,JKPS}.
The explicit expression of the two-vertex in the $\mathfrak{su}(2)$ sector is
\begin{align}
\label{eq:vertexdef}
\langle \rV_2|&=\sum_{\so_1,\ldots,\so_{\LL} =\, \uparrow,\,
\downarrow}\langle \so_1,\ldots,\so_{\LL} |\otimes \langle \so_{\LL}
,\ldots,\so_1|\,\Sigma_2\,.
\end{align}
 Using this formalism we get for the scalar product of two rotated states
\begin{align}
\label{eq:vertexsp}
\langle \rV_2| |\psi_1\rangle|\psi_2\rangle& \equiv \langle
\rV_2|\,\rg_{2}\,|\mathbf{u}_2\rangle\;\rg_{1}\,
|\mathbf{u}_1\rangle\nonumber \\
&=(-1)^{M_2} \langle\,\downarrow^{\LL} \!|
\prod_{j=1}^{M_2} B(u_{2, j}) \,\rg_{2}^{-1}\rg_{1}\,
\prod_{i=1}^{M_1} B(u_{1, i})
 |\!\uparrow^{\LL} \,\rangle\,.
\end{align}
The vertex $\langle \rV_2|$ helps to transfer spin chain operators
from one chain into another following \cite{KKNv,JKPS}:
\begin{align}
\label{eq:reflection}
&\,\langle{\rV}_2|A^{(1)}(u)=\langle{\rV}_2|D^{(2)}(u),\quad
\langle{\rV}_2|B^{(1)}(u)=-\langle{\rV}_2|B^{(2)}(u),\\\nonumber &
\,\langle{\rV}_2|D^{(1)}(u)=\langle{\rV}_2|A^{(2)}(u),\quad
\langle{\rV}_2|C^{(1)}(u)=-\langle{\rV}_2|C^{(2)}(u).
\end{align}\label{eq:reflection2}
It can also be used to transfer the rotations from one chain to another
\begin{align}
&\,\langle{\rV}_2|\,\rg^{(1)}=\langle{\rV}_2|\,(\rg^{-1})^{(2)}\;.
\end{align}
This relation is a consequence of the singlet property of the vertex
$\langle\rV_2|(S^{(1)\alpha}+S^{(2)\alpha})=0$.  Without loss of
generality we can set in equation (\ref{eq:vertexsp})
$\rg=\rg_2^{-1}\rg_1\simeq \rg_z$
\begin{align}
\label{eq:vertexGI}
\langle \rV_2|\,
|\psi_1\rangle|\psi_2\rangle &\simeq\langle\,\downarrow^{\LL}
\!| \prod_{j=1} ^{M_2} B(u_{2, j}) \prod_{i=1}^{M_1} B(u_{1, i})
 \,e^{z
S^-}\,|\!\uparrow^{\LL} \,\rangle \\\nonumber &=\frac{ z^{\LL-M}}{(\LL-M)!}
 \, \langle\,\downarrow^{\LL} \!| \prod_{j=1} ^{M_2} B(u_{2, j})
 \prod_{i=1}^{M_1} B(u_{1, i})
  (S^-)^{{\LL-M}}|\!\uparrow^{\LL} \,\rangle\;,
\end{align}
where in the first line we have neglected a factor which can be
reconstituted from equation (\ref{eq:hws}).  The last line is, up to
the prefactor, the partial domain wall boundary condition partition
function, pDWPF, with $M\equiv M_1+M_2\leq \LL$. The dependence on the global
rotations is through the factor $z^{L-M}$.

%%%%%%%%%%%%%%%%%%%%%%%%%%%%%%%%%
\subsection{Direct and mirror representation for DWPF}
%%%%%%%%%%%%%%%%%%%%%%%%%%%%%%%%%

\begin{figure}
%     \vskip 1cm
         %%----start of first figure----
	 \begin{minipage}[t]{0.4\linewidth}
            \centering
            \includegraphics[width=4.5 cm]{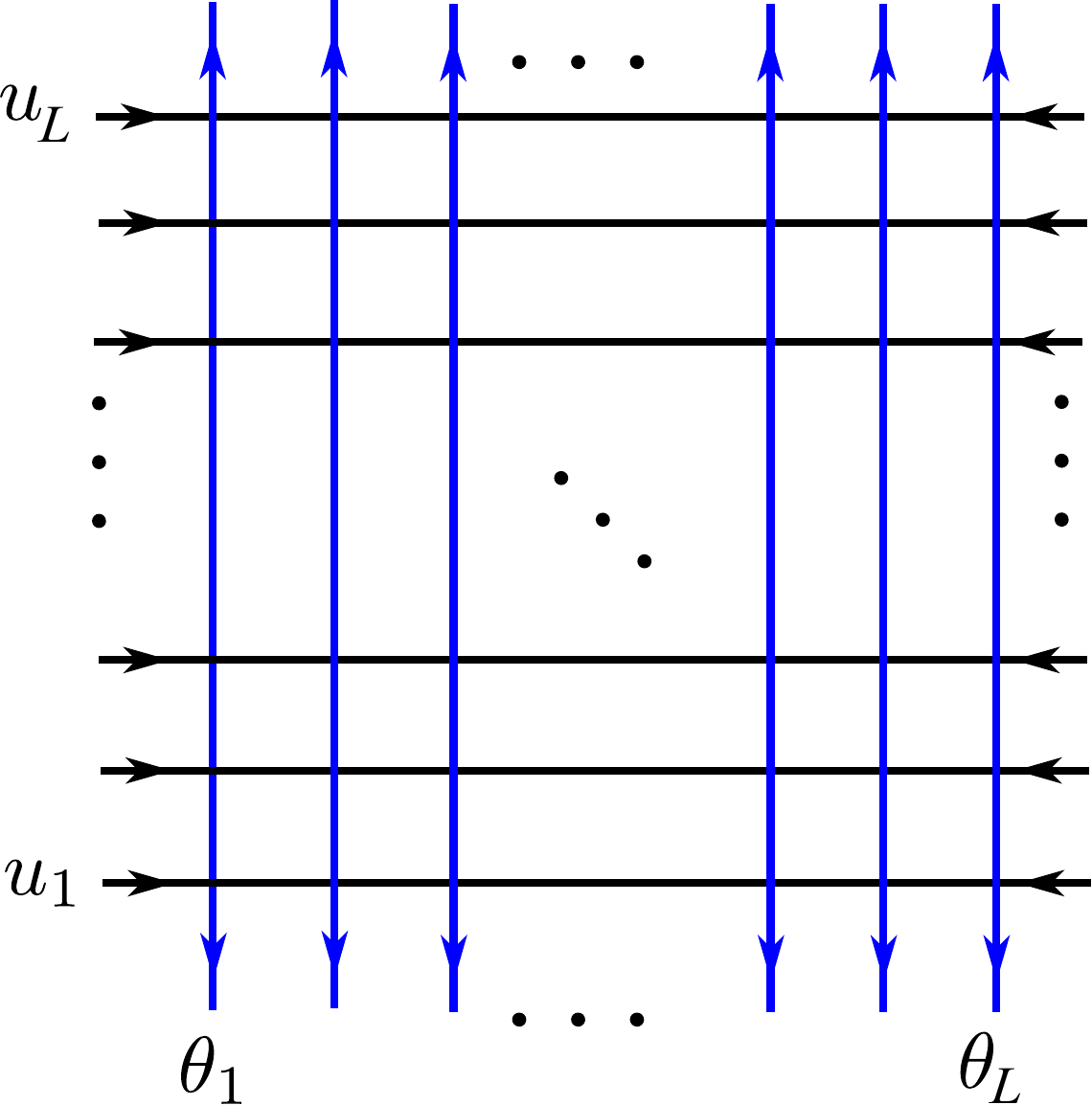}
\caption{The domain wall partition function of six vertex model.
Notice that the configurations at the four boundaries are fixed.}
\label{fig:DWPF}
         \end{minipage}%
         \hspace{1.8cm}%
                \begin{minipage}[t]{0.4\linewidth}
            \centering
            \includegraphics[width=6.3cm]{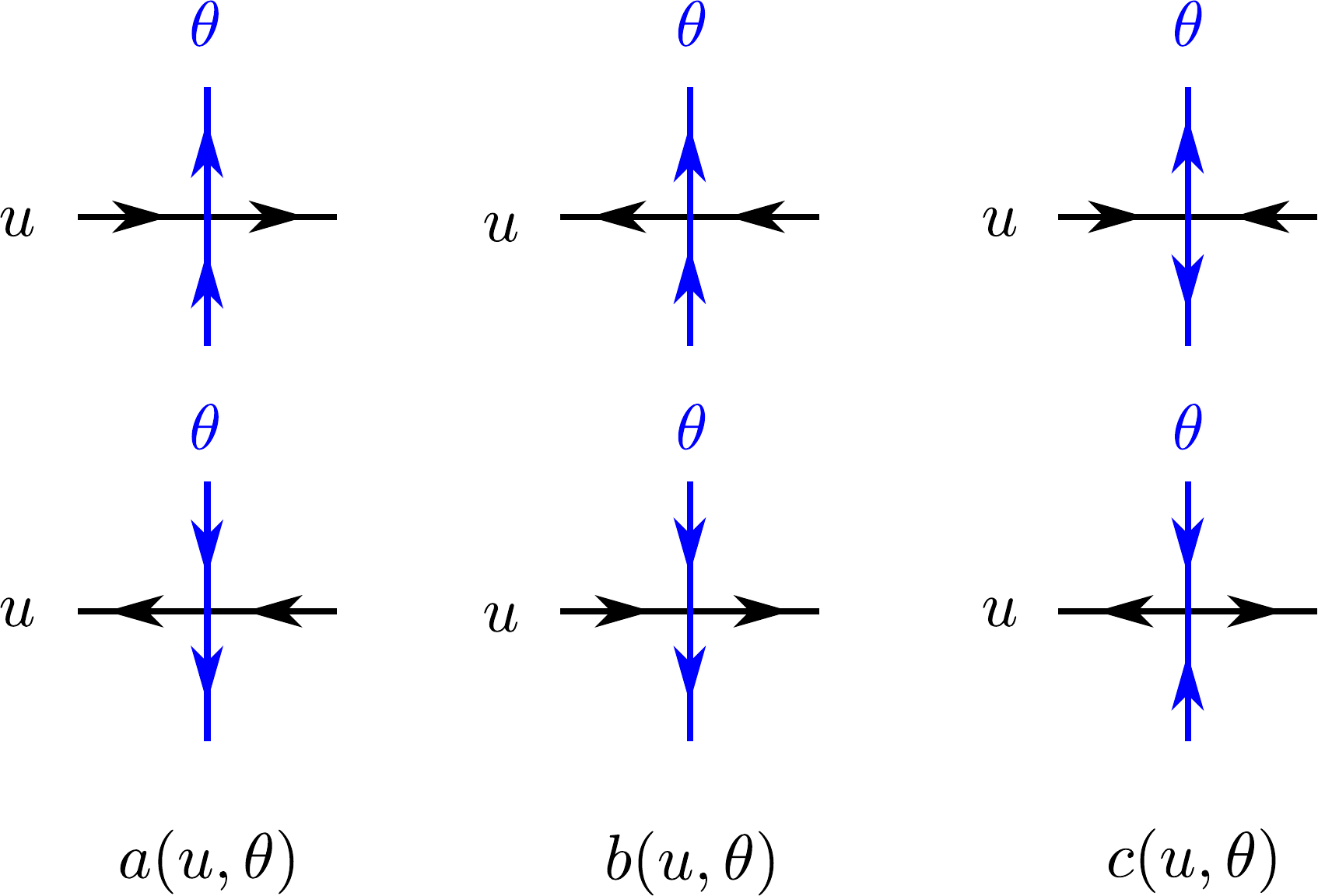}
	    \caption{The six different vertex configuration, with their weights
defined in equation (\ref{sixweights}).}
\label{fig:abc}
         \end{minipage}
      \end{figure}

The domain wall partition function computes the partition
function of the six vertex model on an $L\times L$ grid with domain
wall boundary condition, as is shown in figure \,\ref{fig:DWPF}. \hskip -2mm

The sum in the DWPF concernes all the configurations involving the six
types of vertices shown in figure \ref{fig:abc}, the weight of each
type of vertex being given by
 \begin{align}
 \label{sixweights}
a(u,\theta)=u-\theta+i/2\,, \qquad b(u,\theta)=u-\theta-i/2\,, \qquad
c(u,\theta)=i\,.
\end{align}
The DWPF can be alternatively defined in the language of the algebraic
Bethe Ansatz as
\begin{align}
\label{eq:DWPF}
Z_{\LL} (\uu|\thth )\equiv\langle\,\downarrow^{\LL}
\!|\prod_{k=1}^{\LL} B(u_k;\thth )|\!\uparrow^{\LL} \,\rangle\,,
\end{align}
where $\thth =\{\theta_1,\cdots,\theta_{\LL} \}$ is the set of
inhomogeneities and $\uu=\{u_1,\cdots,u_{\LL} \}$ are the magnon
rapidities.  The mapping from the six vertex configuration to the
algebraic Bethe Ansatz language, explained at length in
\cite{OmarFreezing}, is based on the interpretation of the six
non-trivial vertex configurations (\ref{sixweights}) as the non-zero
elements of the Lax matrix $\rL(u)$.  The black, horizontal lines
correspond to copies of the auxiliary space, while the blue vertical
lines correspond to the quantum space.  Conventionally, the Lax matrix
acts from NE to SW. The DWBC configuration with all the arrows
pointing upwards on the upper edge corresponds to the vacuum
$|\!\uparrow^{\LL} \,\rangle$, the horizontal lines with incoming
arrows correspond to operators $B(u)$ and the horizontal lines with
outgoing arrows correspond to operators $C(u)$.  The symmetry
properties of the Lax matrix are inherited by the six-vertex
configuration,
\begin{align}
 \label{Laxsym}
\rL_n(u)=\rL_n(u)^{t,t_0}=\sigma_n^2\sigma^2\,
\rL_n(u)\,(\sigma_n^2\sigma^2)^{-1}\;,
\end{align}
where $t,t_0$ mean transposition in the quantum and in the auxiliary
spaces respectively.  The simultaneous transposition in the two
spaces, followed by the conjugation with $\sigma^2$ in the two spaces,
amounts to the rotation of the corresponding vertex by $180^\circ$.
The conjugation is necessary to keep the orientation of the arrows
unchanged.

\begin{figure}
     \hskip 0.5cm
         %%----start of first figure----
	 \begin{minipage}[t]{0.9\linewidth}
          \begin{center}
           \includegraphics[scale=0.65]{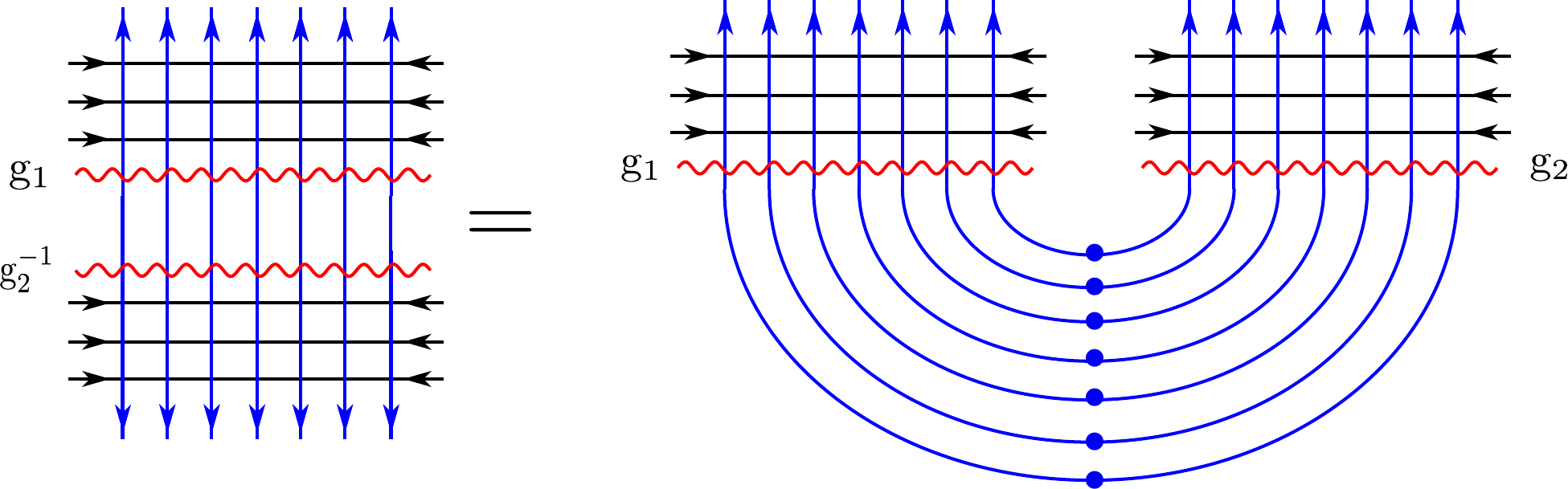}
\caption{Relation between six vertex model configuration and the spin
vertex formalism. The blobs symbolize the action of the operator
$\Sigma_2$:  an extra
factor $(-1)$ if $\so = \uparrow$.}
\label{fig:2v}
    \end{center}             \end{minipage}
      \end{figure}

When applied to the untwisted monodromy matrix, the simultaneous
transposition reverses the order of the sites on the chain, as one can
see as well from the graphical interpretation,
\begin{align}
 \label{monodtransp}
\rT(u)^{t,t_0}=\rL_{\LL} (u)^{t,t_0}\ldots\rL_1(u)^{t,t_0}
=\rL_{\LL} (u)\ldots\rL_1(u)\equiv \bar \rT(u)\;.
\end{align}
In components, this means for example that $B^t(u)=\bar C(u)$, where
the bar denotes the reversal of the order of the sites in the chain.
Taking the transpose in the quantum space and using the fact that the
DWPF is symmetric in the variables $\theta_1,\ldots,\theta_{\LL} $, so
that the order of the sites is irrelevant, one gets
\begin{align}
\label{eq:DWPFt}
Z_{\LL} (\uu|\thth )=\langle\,\uparrow^{\LL} \!|\prod_{k=1}^{\LL}
B^t(u_k;\thth )|\!\downarrow^{\LL} \,\rangle= \langle\,\uparrow^{\LL}
\!|\prod_{k=1}^{\LL} C(u_k;\thth )|\!\downarrow^{\LL} \,\rangle\,.
\end{align}
 The second equality in (\ref{Laxsym}) translates into the following
 equality for the monodromy matrix
\begin{align}
\label{monodsym}
\rT(u)=\Sigma_2 \sigma^2\; \rT(u)\; (\Sigma_2 \sigma^2)^{-1} \;,
\end{align}
with $\Sigma_2$ defined in (\ref{defSigma}).  Written in components, this gives the relation
(\ref{BtoC}) between the $B(u)$ and $C(u)$ operators.
Given that the action of $\Sigma_2$ on the vacua is
$\Sigma_2\,|\!\uparrow^{\LL} \,\rangle=(-1)^{\LL}
\,|\!\downarrow^{\LL} \,\rangle$ and $\Sigma_2\,|\!\downarrow^{\LL}
\,\rangle=|\!\uparrow^{\LL} \,\rangle$ the DWPF can take the
alternative form
\begin{align}
\label{eq:DWPFC}
Z_{\LL} (\mathbf{u}|\thth )=\langle\,\uparrow^{\LL}
\!|\prod_{k=1}^{\LL} C(u_k;\thth )|\!\downarrow^{\LL} \,\rangle\,.
\end{align}
The equality of (\ref{eq:DWPF}), (\ref{eq:DWPFt}) and (\ref{eq:DWPFC})
expresses the invariance of the DWPF under reversal of the arrows.
If one views the action of the monodromy matrix as an evolution in a
(discrete) time, the rotation with $180^\circ$ clockwise can be viewed
as a PT transformation, and the transformations properties above as a
CPT invariance.
It is instructive to go back and interpret equation
(\ref{eq:vertexsp}) in terms of six-vertex configuration.  The second
line can be written in terms of a rectangular six-vertex
configuration.
Now we insert the resolution of identity between the two rotations and
apply transposition and conjugation with $\Sigma_2$ in the first
block,
\begin{align}
\label{eq:vertexsix}
(-1)^{L-M_2 } & \langle\,\downarrow^{\LL} \!| \prod_{j=1}^{M_2}
B(u_{2,j}) \,\rg_{2}^{-1}\rg_{1}\, \prod_{k=1}^{M_1} B(u_{1,k})
|\!\uparrow^{\LL} \,\rangle\,=\\[-4mm] \nonumber \\=\!\!\!\!
\sum_{\so_1,\ldots,\so_{\LL} =\uparrow,\downarrow}& \!\!\!\!\langle
\so_1,\ldots,\so_{\LL} |\,\Sigma_2\, \rg_2\,\prod_{j=1}^{M_2} \bar
B(u_{2,j})|\!\uparrow^{\LL} \,\rangle\, \langle \so_1,\ldots,\so_{\LL}
|\,\rg_{1}\,\prod_{k=1}^{M_1} B(u_{1,k}) |\!\uparrow^{\LL}
\,\rangle\,.\nonumber
\end{align}
In the six-vertex model picture, the first line corresponds to the
ordinary domain wall partition (with the global rotations inserted in
the middle) whereas the second line corresponds to the configuration
where the lower half is rotated by 180$^{\circ}$ (see figure
\ref{fig:2v}).  The action of the operator $\Sigma_2$ is symbolized by
the blobs in the middle of the lines.  The effect of such blob is a
multiplication by a factor $(-1)$ if $\so = \uparrow$.

\begin{figure}[h!]
     \hskip 0.5cm
         %%----start of first figure----
	 \begin{minipage}[t]{0.8\linewidth}
          \begin{center}
           \includegraphics[scale=0.41]{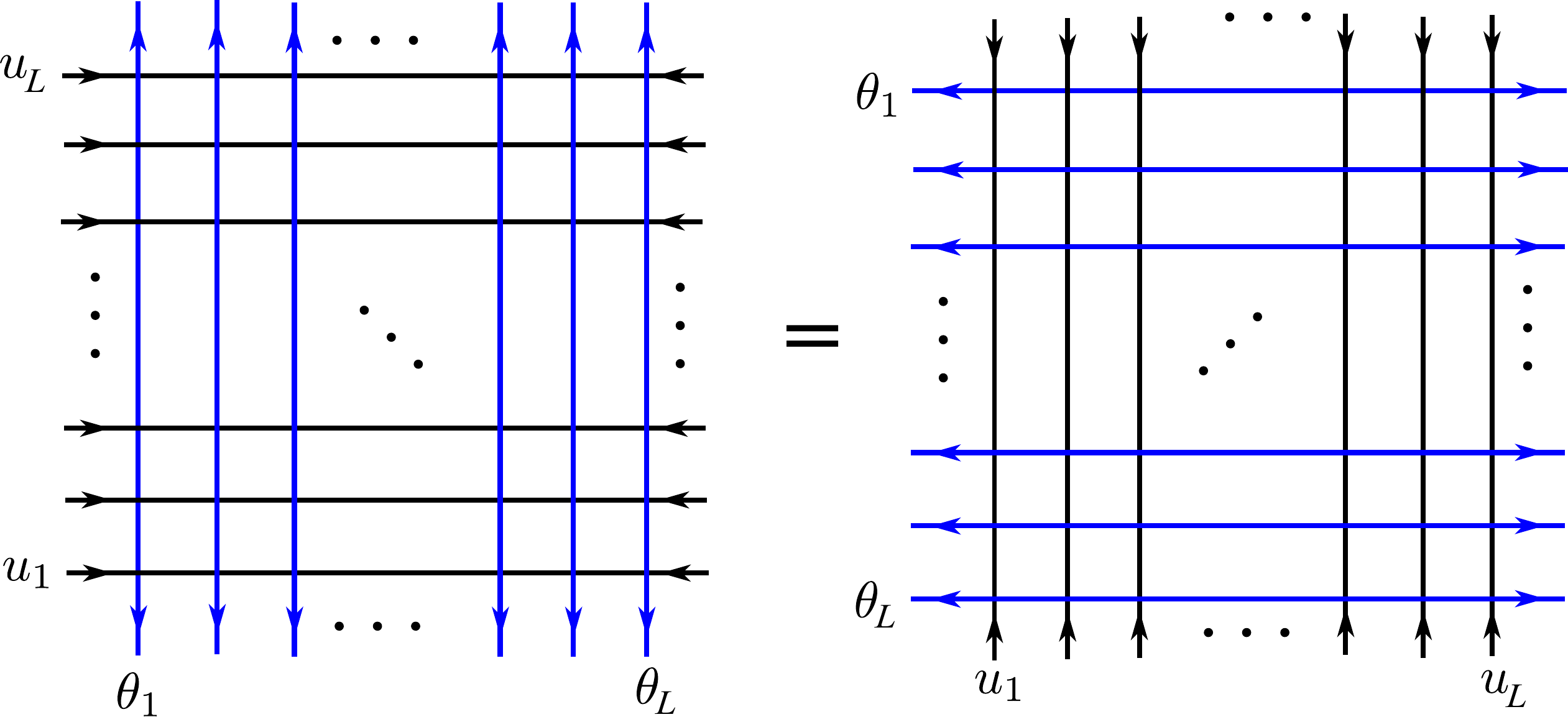}
\caption{We rotate the configuration in figure \,\ref{fig:DWPF} by 90$^\circ$ clockwise, the expression for the partition function is
invariant.  The interpretations of the two diagrams are different.  The
roles of rapidities and inhomogeneities are exchanged in the two
configurations.}
\label{fig:rotateDW}
    \end{center}             \end{minipage}
      \end{figure}

Less obvious are the transformation properties under rotation of the
six-vertex configuration with $90^\circ$.  By the space-time analogy
above we can consider this transformation as an exchange of space and
time, which in the context of integrable field theories goes under the
name of mirror transformation.  We are using this term here, although
there might be differences with other instances of mirror
transformations.  Under the mirror transformation we get another
domain wall boundary condition configuration with the inhomogeneities
and rapidities exchanged as in figure \ref{fig:DWPF}.  The vertices in
figure \ref{fig:abc} transform as follows
\begin{align}
 \label{sixweightsmir}
a(u, \theta)&\to b(\theta,u)=-a(u, \theta)\;,\\ \nonumber
b(u, \theta)&\to a(\theta,u)=-b(u, \theta)\,, \\ \nonumber
c(u, \theta)&\to c(\theta,u)=c(u, \theta)\,.
\end{align}
We conclude that the rotated DWPF shown in figure \ref{fig:rotateDW}
is equal to the non-rotated one up to a sign.  To compute this sign we
need to know the parity of number of vertices of type $a$ and $b$,
$N_a$ and $N_b$.  Due to the particular type of boundary condition we
consider, on each line there should be one type $c$ vertex.  For the
DWPF the sign is then given by
\begin{align}
(-1)^{N_a+N_b}=(-1)^{\LL (\LL-1)}=1
\end{align}
so that
\begin{align}
\label{ZZ}
 Z_{\LL} ({\bm \theta}|\mathbf{u})= Z_{\LL} ({\mathbf u}|{\bm
 \theta})\;.
\end{align}

Therefore, the DWPF in the mirror representation is given by
\begin{align}
\label{BCDWPF}
Z_{\LL} (\thth |\mathbf{u})=\langle\,\uparrow^{\LL}
\!|\prod_{k=1}^{\LL} C(\theta_k;\mathbf{u})|\!\downarrow^{\LL}
\,\rangle=\langle\,\downarrow^{\LL} \!|\prod_{k=1}^{\LL}
B(\theta_k;\mathbf{u})|\!\uparrow^{\LL} \,\rangle.
\end{align}
%

%%%%%%%%%%%%%%%%%%%%%%%%%%%%%%%%%%%
\subsection{Global rotation and twist}
%%%%%%%%%%%%%%%%%%%%%%%%%%%%%%%%%%%%

\begin{figure}[h!]
     \hskip 0.5cm
         %%----start of first figure----
	 \begin{minipage}[t]{0.9\linewidth}
          \begin{center}
           \includegraphics[scale=0.42]{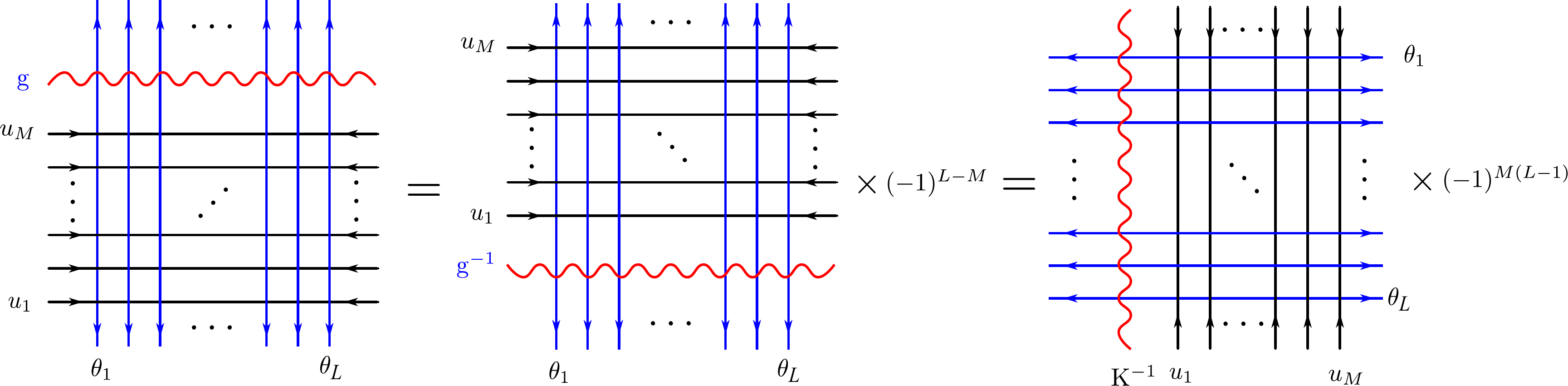}
\caption{The mirror representation in the presence of global rotation.
After turning the diagram 90$^\circ$, the global rotation plays the
role of a twist.}
\label{fig:tDWPF2}
    \end{center}             \end{minipage}
      \end{figure}

The next step is to find the mirror representation of the partition
functions in the six vertex model in presence of a global rotation.
We suppose that, as in equation (\ref{eq:vertexGI}) the rotation acts
directly on the vacuum.  The partition function we study is depicted
in figure \,(\ref{fig:tDWPF2}).  On the l.h.s. the partition function
corresponds to
\begin{align}
\label{lhs}
\texttt{lhs}=\,\langle\uparrow^{\LL} \!|\,\rg^t\,C(u_1)&\cdots
C(u_M)|\!\downarrow^{\LL} \rangle=(-1)^{\LL -M}\langle\downarrow^{\LL}
\!|\,\rg^{- 1}\,B(u_1)\cdots B(u_M)|\!\uparrow^{\LL} \rangle
\nonumber\\
&=\langle\downarrow^{\LL} \!|B(u_1)\cdots B(u_M)\,\rg_{
}\,|\!\uparrow^{\LL} \rangle \,,
\end{align}
where in the last two equalities we have used the conjugation with the
matrix $\Sigma_2$ and transposition in the quantum space respectively,
as discussed in the previous subsection, and $\rg^{-1}=\Sigma_2\,\rg^t
\, \Sigma_2^{-1}$.  After rotation with $90^\circ$ the partition
function acquires a sign $(-1)^{N_a+N_b}=(-1)^{M(L -1)}$, {\it cf.}
equation (\ref{sixweightsmir}), and the rotation in the quantum space
$\rg$ is replaced by the twist matrix $\rK^t$ acting in the auxiliary
space
\begin{align}
\label{rhs}
\texttt{rhs}=(-1)^{M(\LL-1)}\langle\downarrow^M|B_{\rK^t|0}(\theta_1)\cdots
B_{\rK^t|0}(\theta_{\LL} )|\uparrow^M\rangle
\end{align}
We conclude that after the mirror transformation the rotation in the
quantum space $\rg$ is replaced with the twist $\rK^t$.  In particular
\footnote{In what follows, we will denote $\rK^t$ by $\rK$ to avoid
cumbersome notations.}
\begin{align}
\label{mir}
\rg=e^{zS^-}\qquad \buildrel {\rm mirror} \over \longleftrightarrow
\qquad \rK^t=e^{z\sigma^+}\;.
\end{align}
Since in the mirror representation the $B$-operators are twisted, we
can apply the SoV formalism.

%%%%%%%%%%%%%%%%%%%%%%%%%%%%%%%%%%%
\subsection{Two dual integral representations for the scalar product}
%%%%%%%%%%%%%%%%%%%%%%%%%%%%%%%%%%%%

Both the expressions (\ref{lhs}) and (\ref{rhs}) can be written as
multiple integrals using SoV method.  For (\ref{lhs}) the integral
representation was derived in \cite{KKN}, after introducing an extra
twist which can be subsequently set to zero.  Since we are going to
compute scalar products for sub-chains, the rapidities are not assumed
to be on shell.  Nevertheless one can apply the argument of \cite{KKN}
to carry on the computation.

Consider first the {\tt lhs}, eq.  (\ref{lhs}), which we denote using the
same notation as above (but with $|\thth|\ge|\uu|$).  In order to go
to the SoV representation, we introduce a left twist $\rK_{\e|0}$ and
then take the limit $\e\to 0$.  For simplicity we take $\rg=e^{zS^-}$.
Only one term in the expansion of the exponent survives and the result
is
\begin{align}
\texttt{lhs} \, \to \langle\,\downarrow^{\LL} \!|B_\e(u_{1})\ldots
B_\e(u_{M})\,e^{zS^-}|\!\uparrow^{\LL} \,\rangle,
\end{align}
In this way we represent this expectation value as a multiple integral
in the separated variables.  The derivation is  the one from \cite{KKN},
after noticing that only one term in the expansion of
the exponent survives (the one which compensates the extra $S_z$ charge
$L-M$).  We give only the final result,
 \begin{align}
\texttt{lhs} \, &=i^{\LL} {z ^{\LL -M}\over (L -M)!}\ \; \Xi_{\thth }
\oint \limits_{\CC_\thth} \prod_{j=1}^{\LL} {dx_j\over 2\pi i} \
{Q_\uu (x_j) \ e^{2\pi (j-1) x_k}\over Q_\thth^+(x_j)Q_\thth^-(x_j)} \
\prod_{j<k}^{\LL} (x_j-x_k) \ \( \sum _{j=1}^{\LL} x_j\)^{\LL -M}
   \end{align}
where the function $\Xi_{\thth }$ is defined by
\begin{align}
\label{eq:Xi}
\nonumber  \\[-7mm] \Xi_{\thth }=&\,\frac{\prod_{j<k}(\theta_{j}-
\theta_{k})(\theta_{j}- \theta_{k}+i)(\theta_{j}-
\theta_{k}-i)}{\prod_{j<k}\left(e^{2\pi \theta_j}-e^{2\pi \theta_k}
\right)} \\[-7mm] \nonumber
\end{align}
and the contour $\CC_\thth$ encircles the sets $\thth^+ $ and $\thth^-$.
Now we write integral representation for the rhs, eqn.  (\ref{rhs}),
  \begin{eqnarray}
   \begin{aligned}
    \label{Zzz} \texttt{rhs} \, &= (-1)^{M(L -1)} \, i^M \ z^{\LL -M} \
    \Xi_{\mathbf{u}} \ \oint \limits_{\CC_ \uu} \prod_{j=1}^{M}
    {dx_j\over 2\pi i} \ \ \prod_{j<k}^M(x_j-x_k) \ \prod_{k=1}^M
    {Q_\thth(x_k)\ e^{2\pi (k-1) x_k} \over Q_\uu^+(x_k)\
    Q_\uu^-(x_k)} ,\ \ \
  \end{aligned}
\label{rhsSov} \end{eqnarray}
 where $\Xi_{\mathbf{u}}$ is defined similarly as in (\ref{eq:Xi}) and
 the contour $\CC_\uu$ encircles the all points ${\mathbf{u}^+}$ and
 ${\mathbf{u}^-}$.

%%%%%%%%%%%%%%%%%%%%%%%%%%%%%%%%%%%%%%
\section{Three-point functions in the SoV representation}\label{sec:3pt}
\label{3pt}
%%%%%%%%%%%%%%%%%%%%%%%%%%%%%%%%%%%%%%

In this section, we compute the structure constant using the spin
vertex formalism in \cite{JKPS,KKNv,StringBit}.  After the mirror
transformation, we can compute both the spin vertex and the wave
functions of the external states in the SoV representation, and
combining them we get the final result.  We can of course do the
computation with the generic twists, but in fact it is enough to
consider the triangular twists of the following form
\begin{align}
\label{eq:Tritwist}
\rK_{z_a}=e^{z_a\,\sigma^+}=\left(
            \begin{array}{cc}
	      1\ & \ z_a \\
              0\ & \ 1 \\
            \end{array}
          \right),\qquad a=1,2,3.
\end{align}
The reason is that the twists in the mirror representation come from
the global rotations in the original representation.  As is shown in
(\ref{eq:hws}), if we start with on-shell Bethe states which satisfy
highest weight conditions, the most general global rotation can be
reduced to the rotation of the form $\rg_z=e^{z\,S^-}$, together with
some factors which are not relevant.  So we can define our external
states as $|\psi\rangle=e^{z S^-}|\mathbf{u}\rangle$ without loss of
generality.  The global rotation $e^{zS^-}$, upon performing the
mirror transformation, turns into the twist of the form
(\ref{eq:Tritwist}).

%%%%%%%%%%%%%%%%%%%%%%%%%%%%%%%%%
\subsection{Spin vertex and the mirror transformation}
%%%%%%%%%%%%%%%%%%%%%%%%%%%%%%%%%%

We are interested in computing the three-point function for three
operators belonging to the so-called left $\mathfrak{su}(2)_{\LL} $
subsector of the $\mathfrak{so}(4)\simeq
\mathfrak{su}(2)_{_\rL}\otimes \mathfrak{su}(2)_{_\rR}$ sector of the
$\CN=4$ SYM theory.  This sector is made by the scalar fields $X,\
\bar X,\ Z,\ \bar Z$ which belong to the bi-fundamental representation
of $\mathfrak{su}(2)_\rL\otimes \mathfrak{su}(2)_{_\rR}$,
\begin{align}
&|Z\rangle =|\!\uparrow\rangle_{_\rL}\otimes | \!\uparrow
\rangle_{_\rR}\equiv|\!\uparrow\uparrow\rangle\;, \qquad |\bar Z\rangle
=|\!\downarrow\rangle_{_\rL}\otimes |
\!\downarrow\rangle_{_\rR}\equiv|\!\downarrow\downarrow\rangle\;,\\
& |X\rangle = |\!\uparrow \rangle_{_\rL}\otimes |\!
\downarrow\rangle_{_\rR}\equiv|\!\uparrow\downarrow\rangle\;,\qquad
|\bar X\rangle =-|\!\downarrow\rangle_{_\rL}\otimes |
\!\uparrow\rangle_{_\rR}\equiv-|\!\downarrow\uparrow\rangle\;.
\nonumber
\end{align}
Again, we are using the vertex formalism from \cite{KKNv, JKPS,StringBit} to compute the overlap of the three spin chains.
At tree level, the three vertex $\langle \rV_3|$ is composed by the three singlets $_{(ij)}\langle \rV_2|$ corresponding to the  bridges
connecting the piece $(ij)$ of the chain $i$ with the piece $(ji)$ of the chain $j$,
\begin{align}
\langle \rV_3|=\,_{(12)}\langle \rV_2|\otimes \,_{(23)}\langle \rV_2|\otimes \, _{(31)}\langle \rV_2|\,.
\end{align}
We
consider the case where the non-trivial magnon excitations belong only
to the left sector\footnote{ This class of three-point functions are
called type I-I-I or unmixed in \cite{KKNv}.}.  The three external
Bethe states with global rotations $\rg_{z_a}\otimes \rg_{\hat z_a} $
($a=1,2,3$) are given by
\begin{align}
|\Psi_a\rangle&= |\psi_a\rangle_{_\rL}\otimes|\hat\psi_a\rangle_{_R}
\end{align}
with the right components being rotated pseudovacua,
\begin{align}
|\psi_a \rangle_{_\rL}=\rg_{z_a }|\mathbf{u}_a \rangle_{_\rL} , \quad
|\hat \psi_a \rangle_{_\rL}= \hat \rg_{\hat z_a }|\uparrow^{\LL _a
}\rangle_{_\rR} \qquad (a=1,2,3).
\end{align}
The spin vertex  also splits into two identical parts
$\langle\rV_3|={_{_\text{L}}\langle}\rV_3|\otimes{_{_\text{R}}\langle}\rV_3|$.
This insures the complete factorization of the left and right sectors,
\begin{align}
\langle \rV_3| \ |\Psi_1\rangle\,|\Psi_2\rangle\,|\Psi_3\rangle =
\left(\langle \rV_3| \
|\psi_1\rangle\,|\psi_2\rangle\,|\psi_3\rangle\right)_{_\rL}\;\left(\langle
\rV_3|\,\hat \rg_{\hat z_1}\,|\uparrow^{\LL _1}\rangle\,\hat\rg_{\hat
z_2}\,|\uparrow^{\LL _2}\rangle\,\hat\rg_{\hat z_3}|\uparrow^{\LL
_3}\rangle\right)_{_\rR}.
\end{align}
The right piece is  easily calculted and is equal to
$(\hat z_{12})^{L_{12} }(\hat z_{23})^{L_{23}} (\hat z_{31})^{L_{31} }$, where $L_{ab}=\frac{1}{2}(L_a+L_b-L_c)$, $(a,b,c=1,2,3)$.

\begin{figure}[h!]
 \vskip 0.5cm    \hskip 0.4cm
         %%----start of first figure----
	 \begin{minipage}[t]{0.97\linewidth}
          \begin{center}
           \includegraphics[scale=0.37]{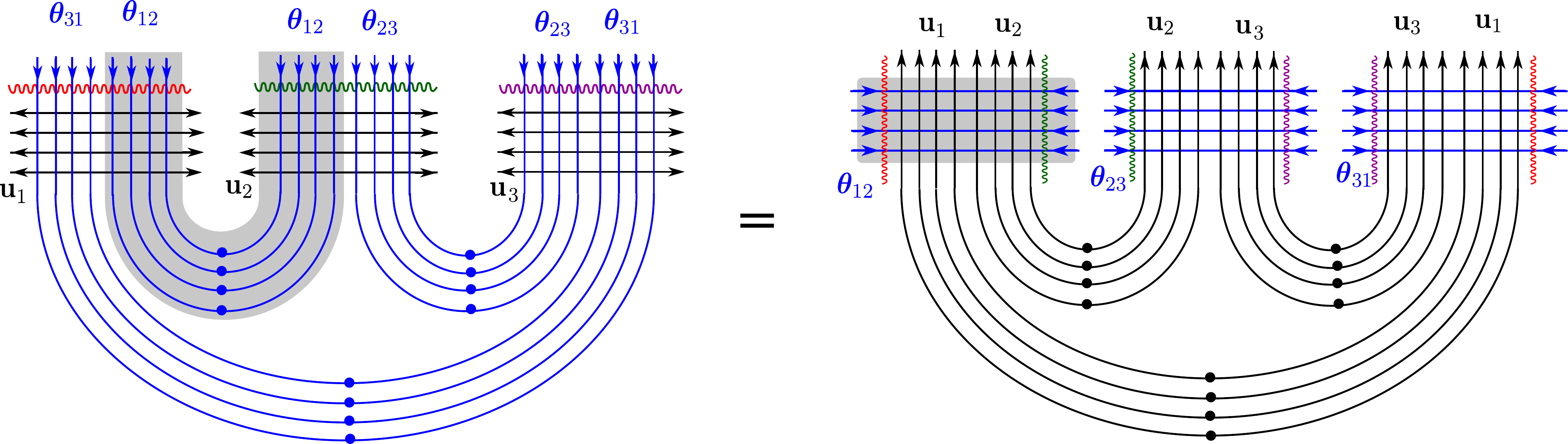}
\caption{The three-point function before and after the mirror
transformation.  One can see that the role of impurities and
rapidities are exchanged, as in the case of scalar products.  In
addition, the global rotation becomes twist in the mirror
representation, as in the scalar product.}
\label{fig:mirror3}
    \end{center}             \end{minipage}
      \end{figure}

As mentioned before, we take the following external states
\begin{align}
|\psi_a \rangle=&\,e^{z_a S^-_a }|\mathbf{u}_a \rangle=
\prod_{j=1}^{M_a} B(u_{a,j}) \ e^{z_a S^-_a }\,|\uparrow^{\LL _a
}\rangle,\qquad a=1,2,3.
\end{align}
We are going to concentrate from now on on the structure constant in
the left sector and drop the L index on the states and on the vertex
\begin{align}
C_{123}^{\rL}=\langle\rV_3|\, |\psi_1\rangle_{_\text{L}}
|\psi_2\rangle_{_\text{L}}|\psi_3\rangle_{_\text{L}}.
\end{align}
The structure constant above has a representation in terms of 6-vertex
model partition function on the diagram of figure \ref{fig:mirror3}.
Apart from the blobs, this diagram is nothing but a redrawing of the
hexagon DWPF in figure \ref{fig:conic} (in the hexagon DWPF there was
no need to put blobs along the ``bridges").

We will perform a mirror transformation on the six-vertex
configuration of the three-point function as is shown in figure
\ref{fig:mirror3}.  We rotate the left subchains clockwise and the
right subchains anti-clockwise and combine together the neighboring
subchains, as is shown in the shaded region of figure
\ref{fig:mirror3}.  Potentially, there is some minus sign coming
from the transition from the direct vertex $\langle\rV_3|$ to the
mirror vertex $\langle\tilde{\rV}_3|$.  But since the signs coming
from ``bridges'' in the direct and the mirror vertex are related,
since ones constitute boundary conditions for the others, we neglect
any overall sign which may occur.  From the 6-vertex
configuration, we see that we need to compute the following mirror
structure constant \footnote{Here ``$=$'' means equal up to some
overall minus sign.}
\begin{align}
\label{eq:Structure}
{C}_{123}^{\rL} =\tilde{C}_{123}^{\rL} \equiv \langle\tilde{\rV}_3|
|\tilde{\psi}^{(12)}\rangle |\tilde{\psi}^{(23)} \rangle \,
|\tilde{\psi}^{(31) }\rangle.
\end{align}
In what follows, we put a tilde on the operators which are in the
mirror representation.  The three mirror external states are now
\begin{align}
\label{eq:externalstates}
|{\tilde\psi^{(ab)}}\rangle=&\,\prod_{k=1}^{\LL
_{ab}}\tilde{B}_{z_a|-z_{b}}(\theta^{(ab)}_k)\,
|\!\!\uparrow^{M_a+M_{b}}\rangle, \qquad (ab) = (12), (23), (31)\,,
\end{align}
where the operators $\tilde{B}_{z_a|-z_{b}}(\theta )$ are the matrix
elements (first row and the second column) of the following double
twisted monodromy matrices
\begin{align}
\tilde{\rT}_{z_a|-z_{b}}(\th)=&\,\rK_{z_a}\,\prod_{j=1}^{M_a}\rL_{n}
(\th-u_{a,j})\prod_{k=1}^{M_{b}}\rL_k(\th-u_{b,k})\,\rK_{-z_{b}}.
\end{align}
We can construct the SoV states for the three double twisted spin
chains, which are denoted as $|{\xx }\rangle_{z_a|-z_{b}}$.

%%%%%%%%%%%%%%%%%%%%%%%%%%%%%%%%%%%%%%%%%
\subsection{Spin vertex in SoV representation}
%%%%%%%%%%%%%%%%%%%%%%%%%%%%%%%%%%%%%%%%%

In order to compute the spin vertex, we apply the important property
(\ref{eq:reflection}).  In the presence of the twists, this property
is modified to be
\begin{align}
\label{eq:doubleRef}
&\,\langle{\rV}_2|A^{(1)}_{z_1|-z_2}(u)=\langle{\rV}_2|D^{(2)}_{z_2|-z_1}(u),\quad
\langle{\rV}_2|B^{(1)}_{z_1|-z_2}(u)=-\langle{\rV}_2|B^{(2)}_{z_2|-z_1}(u),\\\nonumber
&\,\langle{\rV}_2|D^{(1)}_{z_1|-z_2}(u)=\langle{\rV}_2|A^{(2)}_{z_2|-z_1}(u),\quad
\langle{\rV}_2|C^{(1)}_{z_1|-z_2}(u)=-\langle{\rV}_2|C^{(2)}_{z_2|-z_1}(u).
\end{align}
By putting one of the twists to zero, we obtain similar relations for
the left or right twisted monodromy matrices.  Notice that the left
twisted monodromy matrix is translated to a right twisted one by the
spin vertex and vice versa.  Using these relations, one can show that
\begin{align}
\langle\rV_2|\, |\xx
\rangle_{z|0}|\yy\rangle_{0|-z}={_{z|0}\langle}\yy|\xx
\rangle_{z|0}=(\mu_{z|0}(\xx ))^{-1}\delta_{\xx ,\yy}
\label{eq:actV2}\end{align}
where
%
%\begin{align}
%\label{eq:actV2}
$\delta_{\xx ,\yy}=\delta_{x_1,y_1}\cdots\delta_{x_N,y_N}$
%\end{align}
and $\mu_{z|0}(\xx )$ is the Sklyanin measure.\par

 We can write the three-point spin vertex in the SoV using the
 resolution of identities of the SoV basis.  Denoting by $\xx^{(ab)} $
 are the SoV variables associated with the state $\tilde \psi^{(ab)}$,
 we write the 3-vertex as a triple sum
\begin{align}
\label{eq:V3SoV}
\langle\tilde{\rV}_3|=\,\sum_{\xx ^{(12)},\xx
^{(23)},\xx^{(31)}}\tilde{\rV}(\xx ^{(12)},\xx ^{(23)},\xx^{(31)}) &\,
\mu_{z_1|-z_2}(\xx
^{(12)})\,\mu_{z_2|-z_3}(\xx^{(23)})\,\mu_{z_3|-z_1}(\xx
^{(31)})\\\nonumber &\,\times{_{z_1|-z_2}\langle\xx ^{(12)}|} \otimes
{_{z_2|-z_3}\langle\xx ^{(23)}|} \otimes {_{z_3|-z_1}\langle\xx
^{(31)}|},
\end{align}
where the coefficient function $\tilde{\rV}(\xx ^{(ab)})$ is given by
\begin{align}
\label{eq:VertexSoV}
\tilde{\rV}(\xx ^{(12)},\xx
^{(23)},\xx^{(31)})=\langle\tilde{\rV}_3|\, |\xx
^{(12)}\rangle_{z_1|-z_2}\otimes |\xx ^{(23)}\rangle_{z_2|-z_3}\otimes
|\xx ^{(31)}\rangle_{z_3|-z_1}.
\end{align}
In order to compute this function, we first split the SoV states as is
described in section\,\ref{sec:SoVresult} appendix \ref{AppendixC},
\begin{align}
|\xx ^{(12)}\rangle_{z_1|-z_{2}}=\sum_{\yy_1,\,
\yy_{2}}\mu_{z_1|0}(\yy_1)\ \mu_{0|-z_{2}}(\yy_{2})\
\Phi(\yy_1;\yy_{2}|\xx ^{(12)})\
\,|\yy_1\rangle_{z_1|0}\otimes|\yy_{2}\rangle_{0|-z_{2}},
\end{align}
and similarly for $|\xx ^{(23)}\rangle_{z_2|-z_{3}}$ and $|\xx
^{(31)}\rangle_{z_3|-z_{1}}$.  Acting the states on the three-point
spin vertex and using (\ref{eq:actV2}), we obtain
\begin{align}
\label{eq:SoVkernal}
\tilde{\rV}(\xx ^{(12)},\xx
^{(23)},\xx^{(31)})=&\,\sum_{\yy_1,\yy_2,\yy_3}\mu_{z_1|0}(\yy_1)\
\mu_{z_2|0}(\yy_2)\ \mu_{z_3|0}(\yy_3)\\\nonumber &\,\times
\Phi(\yy_1;\yy_2|\xx ^{(12)})\ \Phi(\yy_2;\yy_3|\xx ^{(23)})\
\Phi(\yy_3;\yy_1|\xx ^{(31)}).
\end{align}
The eigenvalues for the three sets of separated variables of the
subchains are
\begin{align}
\yy_1=\mathbf{u}_1\pm \hfi
 ,\quad
\yy_2=\mathbf{u}_2\pm\hfi,\quad
\yy_3=\mathbf{u}_3\pm\hfi\, ,
\end{align}
where $\yy_a=\mathbf{u}_a\pm\hfi$ is understood as
$y_{a,k}=u_{a,k}\pm \hfi$, ($k=1,\cdots,M_a$).

%%%%%%%%%%%%%%%%%%%%%%%%%%%%%%%%%%%%%%%%%
\subsection{The wave functions}
%%%%%%%%%%%%%%%%%%%%%%%%%%%%%%%%%%%%%%%%%

The external states in the mirror representation are given in
(\ref{eq:externalstates}).  Acting these states on the spin vertex
(\ref{eq:V3SoV}) and using the property of the SoV basis
\begin{align}
_{z_1|-z_2}\langle\xx |\tilde{B}_{z_1|-z_2}(\th)=(z_1-z_2)\,Q_{\xx
}(\th)\,{_{z_1|-z_2}\langle\xx |},
\end{align}
we obtain the wave functions of the mirror Bethe states
\begin{align}
\label{eq:wavefunction}
_{z_a|-z_{b}}\langle\xx
^{(ab)}|\tilde{\psi}^{(ab)}\rangle=& \,(-1)^{\LL
_{a,b}(M_a+M_{b})} \,(z_a-z_{b})^{\LL
_{ab}}\,f_{z_a|-z_{b}}(\xx ^{(ab)})\prod_{k=1}^{M_a+M_{b}}Q_{\thth
^{(ab)}}(x_k^{(ab)})\, ,
\end{align}
where the phase factor $(-1)^{\LL _{ab}(M_a+M_b)}$ comes from the
rewriting of products of $Q$-functions
\begin{align}
\prod_{j=1}^{\LL _{ab}}Q_{\xx _a}(\theta^{(ab)}_j) \equiv
(\thth^{(ab)} - \xx^{(ab)} )=(-1)^{\LL
_{ab}(M_a+M_b)}\prod_{k=1}^{M_a+M_b}Q_{\thth ^{(ab)}}(x^{(ab)}).
\end{align}
The functions $f_{z_a|-z_b}(\xx )$ are the projection of SoV basis on
the pseudovacuum and are given by
\begin{align}
f_{z_a|-z_{b}}(\xx ^{(ab)})=&\,{_{z_a|-z_{b}}\langle\xx ^{(ab)}|}\!
\!\uparrow^{M_a+M_{b}}\rangle.
\end{align}
From (\ref{eq:vacP}) the projection of SoV basis on the pseudovacuum
for the triangular twists of the form (\ref{eq:Tritwist}) is simply
\begin{align}
{_{z_a|-z_{b}}\langle}\xx |\!\uparrow^{\LL} \,\rangle=1.
\end{align}

%%%%%%%%%%%%%%%%%%%%%%%%%%%%%%%%%%%%%%%%%
\subsection{The final result}
%%%%%%%%%%%%%%%%%%%%%%%%%%%%%%%%%%%%%%%%%

We can now assemble the results from the previous subsections and
write down the final result for the structure constant.  Plugging
(\ref{eq:V3SoV}), (\ref{eq:VertexSoV}), (\ref{eq:SoVkernal}) and
(\ref{eq:wavefunction}) into (\ref{eq:Structure}), we obtain
\begin{align}
\label{eq:C123sum}
C_{123}^{\rL}=& \prod_{a=1}^3 (-1)^{L_aM_a} \prod_{(ab)} z_{ab}^{\LL
_{ab}} \ \sum_{\{ \xx ^{(ab)} \}}\sum_{\{ \yy_a \} } \nonumber \\
 \times &\ \prod_{a=1}^3 \mu_{z_a|0}(\yy_a) \ \prod_{(ab)}
 \mu_{z_a|-z_b}(\xx ^{(ab)})\ \ \Phi(\yy_a;\yy_b|\xx ^{(ab)})\ \
 \prod_{k=1}^{M_a+M_b}Q_{\thth ^{(ab)}}(x^{(ab)}_{k}) ,
\end{align}
where we have defined $z_{ab}=z_a-z_b$.  In (\ref{eq:C123sum}), the
summation is over all the possible eigenvalues of all the SoV
variables.  The structure of the summand is the product of Sklyanin
measures of the three spin chains and three subchains, the three
splitting functions which originate from cutting the spin chains, and
the three wave functions.\par

Similar to the scalar product, there's another way of writing
$C_{123}^{\rL}$ which converts the sum over eigenvalues of separated
variables to a multiple contour integral.  The derivation of the
integral representation is analogues to the scalar product in
\cite{KKN}.  The result is given by the integral
\begin{align}
\nonumber C_{123}^{\rL}=\,\texttt{factor} \times\oint
\limits_{\{\uu_{1,2,3} \pm i/2\}} \prod_{(ab)} &
\frac{\Gamma\left(i(\mathbf{u}_a^+-\mathbf{u}_{b}^-)\right)}
{\Gamma\left(i(\yy_a-\yy_{b})\right)}\
\frac{\Gamma\left(1-i(\mathbf{u}_a^+-\xx
^{(ab)})\right)\Gamma\left(1+i(\mathbf{u}_{b}^--\xx ^{(ab)})\right)}
{\Gamma\left(1-i(\yy_{a}-\xx
^{(ab)})\right)\Gamma\left(1+i(\yy_{b}-\xx ^{(ab)})\right)} \\
\label{finalresult} &\times (\xx^{(ab)} - \thth^{(ab)} )\
T(z_1,z_2,z_3)\ d\mu(\xx ^{(ab)} ) \ d\mu(\yy_a ) ,
 \end{align}
 where $(ab)= (12), (23), (31)$.  We denoted by $T(z_1,z_2,z_3)$ the
 factor in the integrand which depends on the twists:
\begin{align}
\label{defTfac}
T(z_1,z_2,z_3)&= \prod_{(ab)} \( {iz_{ab}\over z_a}\)^{N_+^{\yy_b}} \(
{z_{ab}\over iz_b}\)^{N_- ^{\yy_a}} (- z_a) ^{N_+^{ \xx ^{(ab)} }}
(z_b )^{N_-^{\xx ^{(ab)} }}
\end{align}
with
\begin{eqnarray}
 N^{\yy_a}_{\pm} \equiv   {M_a\over 2} \mp i
\sum_{j=1}^{M_a} (y_{a,j} - u_{a,j}), \ \
N^\xx_\pm = {M_a+M_b\over 2} \mp i\!\!\!\sum_{j=1}^{M_a+M_b}
 (x^{(ab)}_j - u_{a,j}-u_{b,j}) . \ \
 \end{eqnarray}
\bigskip
The measures $d\mu(\xx ^{(ab)})$ and $\mu(\yy_a)$ for the spin chains
and subchains are defined by
\begin{align}
\label{defmeasuremu} d\mu(\xx ^{(ab)})=&\,\prod_{k=1}^{M_a+M_{b}} \frac{d
x^{(ab)}_{k}}{2\pi i} \frac{\Delta (\xx^{(ab)} )\Delta(e^{2\pi
\xx^{(ab)} })} { (\xx^{(ab)}- \uu_a^+)(\xx^{(ab)}-
\uu_a^-)(\xx^{(ab)}-\uu^+_b)(\xx^{(ab)}-\uu^-_b)} ,\\ \nonumber
\\[-5mm] d\mu(\yy_a)=&\,\prod_{j=1}^{M_a}\frac{dy_{a,j}}{2\pi i}\,
\frac {\Delta(\yy_{a}) \Delta (e^{2\pi \yy_{a}})}{ (\yy^a -
\uu^+_a)(\yy_a-\uu^-_a) }.  \label{defmeasureom}\end{align}
Finally, the overall factor reads, up to a phase factor,
\begin{align}
\texttt{factor}=&\,
\frac{1}{(M_1+M_2)!(M_2+M_3)!(M_3+M_1)!\,M_1!M_2!M_3!}\\\nonumber
\\[-5mm]\nonumber &\,\times \prod_{(ab)} (z_{ab})^{L_{ab} -M_a-M_b }
\prod_a (z_a)^{-M_a} \\\nonumber \\[-5mm]\nonumber
&\,\times\Xi_{\mathbf{u}_1\cup\mathbf{u}_2}\times\Xi_{\mathbf{u}_2\cup\mathbf{u}_3}
\times\Xi_{\mathbf{u}_3\cup\mathbf{u}_1}\times
\Xi_{\mathbf{u}_1}\times \Xi_{\mathbf{u}_2}\times \Xi_{\mathbf{u}_3}.
\end{align}
The \texttt{factor} comes from the product of the common factor in
(\ref{eq:C123sum}) and the part of Sklyanin measures which do not
depend on SoV variables, Eq.  (\ref{eq:Xi}).

As proved in \cite{KKNv}, the $z_a$ dependence is completely
determined as follows by the Ward identity when operators are primary:
\begin{align}\label{fromward}
C_{123}^{\LL} \ \propto \ (z_{12})^{\LL _{12}-M_1-M_2 +M_3 } \
(z_{23})^{\LL _{23}-M_2-M_3 +M_1 } \ (z_{31})^{\LL _{31}-M_3-M_1 +M_2
}\,.
\end{align}
One way to see this $z$-factorisation explicitly is to first compute
the integral, use the Bethe equations and eliminate $z$'s\fn{Such
manipulation was performed in appendix L of \cite{BKV} in order to
compare the predictions from the hexagon vertex with the weak coupling
result.}.  Although we haven't succeeded, it would be much more
desirable if we could rewrite the integrand (using Bethe equations or
Baxter equations) in such a way that the $z$-independence becomes
manifest.  We leave this as an important future problem.

%%%%%%%%%%%%%%%%%%%%%%%%%%%%%%%%%%%%%%%%%
\section{Conclusion and prospects}
%%%%%%%%%%%%%%%%%%%%%%%%%%%%%%%%%%%%%%%%%

In this paper, we derived a new integral expression for three-point
functions in the $\mathfrak{su}(2)$ sector of $\mathcal{N}=4$ SYM using Sklyanin's
separation of variables.  In order to apply the SoV method, we first
mapped the three-point function to the partition function of the
six-vertex model with a hexagonal boundary and then performed
90$^{\circ}$ rotations, which we call mirror rotations.  The SoV
approach can be readily used after this manipulation without the need
of introducing boundary twists.  The intriguing feature of our result
is that the rapidities (and the extra  inhomogeneities) enter only through
the Baxter polynomials, which are considered to be intimately related
to the quantum spectral curve approach \cite{QS}.  In this sense, the
result obtained in this paper may be regarded as a (small) step toward
the quantum-spectral-curve approach to the structure constants.

We focused only on the $\mathfrak{su}(2)$ sector in this paper.  It would be an
interesting future problem to extend the analysis performed here to
other sectors.  Of particular interest would be the generalization to
the $\mathfrak{sl}(2)$ sector, where the SoV basis already exists \cite{sl2}.  In
the $\mathfrak{sl}(2)$ spin chain, the quantum space and the auxiliary space belong
to different representations in the conventional formulation.  Thus,
the mirror rotation, if it exists, will take a very different form.
For the scalar products, it is possible to write down an integral
expression\footnote{The ordinary SoV integral expression for the
scalar products in the $\mathfrak{sl}(2)$ sector is already known in the literature
\cite{sl2,Sobko}.} akin to the mirror representation in the $\mathfrak{su}(2)$
sector \cite{Komatsu}.  It would be worth investigating if a similar
expression can be obtained also for the three-point function itself.

Another interesting future direction is to understand the relation
between the mirror rotation employed in this paper and the ``genuine''
mirror transformation used in the worldsheet S-matrix approaches
\cite{crossing,mirror}.  At the moment, it is not clear (at least to
us) whether such a connection exists at all.  However, it would be
very intriguing if we could establish the relation as it may pave
the way toward understanding the string world-sheet theory from the
perturbative gauge theory.  Also for this purpose, studying other
sectors will be useful\footnote{To address such a question, it might
be helpful to formulate the mirror rotation used in this paper as some
kind of the (anti-)automorphism of the underlying algebra, as was the
case for the crossing and mirror transformations of the worldsheet
S-matrix \cite{crossing,mirror}.}.

In order to see if our expression has a neat finite-coupling analogue,
it would be important to study in detail how the expression would be
modified at one loop.  To this end, it would be helpful to apply the
approaches developed in \cite{tailoring4,Fix!}, where it was shown
that a certain class of the one-loop structure constants can be
obtained from the leading order result by a clever use of the
inhomogeneities.  It would also be interesting to try to factor out
the dependence on the angles $z_{ab}$ in (\ref{finalresult}) using the
Bethe (or Baxter) equation and further simplify the final expression.

In deriving the integral expression, we obtained several new results
about the SoV basis, such as the explicit expression of the states in
the presence of twists and the splitting function which determines the
overlap between the SoV state in the original chain and the SoV states
in the subchains.  These results may be useful in other problems, such
as the computation of the form factors \cite{Andrei is a nice guy,
Kitanine1, Maillet, Kitanine2} and the entanglement entropy
\cite{entanglement} in the spin chain.  It would also be interesting
if we could use such results to study the hexagon vertex in the SoV
basis.  This may give some hints about how to incorporate the quantum
spectral curve techniques into the hexagon-vertex framework.

We hope that the materials studied in this paper will play a
foundational role in the future progress and help unravelling still
enigmatic features of the gauge/string duality.

\section*{Acknowledgements} Research at the Perimeter Institute is
supported by the Government of Canada through NSERC and by the
Province of Ontario through MRI. D.S., S.K and I.K. gratefully
acknowledge support from the Simons Center for Geometry and Physics,
Stony Brook University.  The research of Y.J, I.K. and D.S. leading to these
results has received funding from the European Union Seventh Framework
Programme FP7-People-2010-IRSES under grant agreement no 269217 and
from  the People
Programme (Marie Curie Actions) of the European Union's Seventh
Framework Programme FP7/2007-2013/ under REA Grant Agreement No 317089.
 I.K, S.K and D.S. thank FAPESP grant 2011/11973-4 for funding their visits to ICTP-SAIFR where part of this work was done.

\appendix
%%%%%%%%%%%%%%%%%%%%%%%%%%%%%%%%%%
\section{The Sklyanin measure}
 \label{AppendixA}
 %%%%%%%%%%%%%%%%%%%%%%%%%%%%%%%%%%%
Using the explicit representations of the SoV basis derived in
(\ref{eq:SoVleft}), let us determine the Sklyanin measure.  Here and
throughout the appendices we consider general $\mathfrak{sl}(2)$ left
and right twists with matrices $\rK_1$ and $\rK_2$ with the following
notations,
\begin{align}
\rK_i=\left(
                                     \begin{array}{cc}
                                       a_i \ &\   b_i\\
                                       c_i \ & \  d_i\\
                                     \end{array}
                                   \right)\;, \qquad
                                   \rK_1\rK_2=\left(
                                     \begin{array}{cc}
                                       a_1a_2+ b_1c_2\ &\   a_1b_2+b_1d_2\\
				       c_1a_2+d_1c_2 \ & \
				       c_1b_2+d_1d_2\\
                                     \end{array}
                                   \right)\equiv \left(
                                     \begin{array}{cc}
                                       a_{12} \ &\   b_{12}\\
                                       c_{12} \ & \  d_{12}\\
                                     \end{array}
                                   \right)\;.
\end{align}
 The Sklyanin measure is given by the inverse of the norm of the SoV
 state as follows
\begin{align}
\langle\xx ^{\prime}|\xx \rangle=\mu_{\rK_1|\rK_2}^{-1}(\xx
)\;\delta_{\xx ^{\prime},\xx }\period
\end{align}
 In \cite{KKN}, the measure was shown to satisfy a certain difference
 equation and was determined up to the overall constant by solving the
 difference equation.  However, using the representations of the SoV
 basis \eqref{ltob} and \eqref{eq:SoVleft}, one can derive the
 following more explicit formula
\begin{align}
 \mu_{\rK_1|\rK_2}^{-1}\left(\theta_1 +\hfi  s_1, \ldots,\theta_{\LL} +\hfi
 s_{\LL}  \right)=\bra{\uparrow^{\LL} }\prod_k \left[
 \frac{A_{\rK_1\rK_2|0}(\theta_k-\hfi s_k  )}{Q_{\thth }(\theta_k-i s_k
 )}\right]\ket{\!\downarrow^{\LL} }\label{mu1}
\end{align}
By decomposing $A_{\rK_1\rK_2|0}$ in terms of the $A$, $B$, $C$ and
$D$ operators for the untwisted monodromy and using the conservation
of the $\mathfrak{su}(2)$ spin, the right hand side of \eqref{mu1} can be further
simplified as follows
 \begin{align}
 \mu_{\rK_1|\rK_2}^{-1}\left(\theta_1 +\hfi s_1, \ldots,\theta_{\LL} +\hfi
 s_{\LL} \right)=(b_{12})^{\LL} \;\bra{\uparrow^{\LL}
 }\prod_{k}\left[\frac{C(\theta_k -\hfi s_k )}{Q_{\thth }(\theta_k-i s_k
 )}\right]\ket{\!\downarrow^{\LL} }\,\label{mu2}\,.
\end{align}
 Note that the right hand side of \eqref{mu2} is nothing but the
 domain wall partition function\fn{Indeed, using the $\mathfrak{su}(2)$ symmetry,
 one can write the right hand side of \eqref{mu2} alternatively as
\begin{align}
( b_{12})^{\LL} \bra{\downarrow^{\LL} }\prod_{k}\left[\frac{B(\theta_k
-\hfi s_k )}{Q_{\thth }(\theta_k-is_k )}\right]\ket{\!\uparrow^{\LL}
}\period
 \end{align}
 }  Since the measure factor is known to satisfy the difference
 equation, in order to determine it unambiguously, it is enough to
 calculate it at one particular value.  The simpletst one to compute
 is $\mu_{\rK_1|\rK_2}^{-1}(\theta_1-\hfi,\ldots , \theta_{\LL} -\hfi)$
 and the result is given as follows:
\begin{align}
 \mu_{\rK_1|\rK_2}^{-1}(\theta_1^-,\ldots , \theta_{\LL} ^-) =
 (b_{12})^{\LL} \period
 \end{align}
 Then the Sklyanin measure can be determined as $(\th_{jk} = \th_j-\th_k)$
\begin{align}
 \mu_{\rK_1|\rK_2}(\xx )={(b_{12})^{-L}
 }\;{\prod_{j<k}(\theta_{jk})(\theta_{jk}+i)(\theta_{jk}-i)}\;{\rm
 Res}_{\yy=\xx }\left[\frac{\prod_{j<k}(y_j-y_k)}{\prod_n Q_{\thth
 }^{+}(y_n)Q_{\thth }^{-}(y_n)}\right]\period
\end{align}
In fact, this measure factor correctly reproduces the result given in
\cite{KKN}, in which the overall normalization factor was determined
by comparison with the known formulas.

 %%%%%%%%%%%%%%%%%%%%%%%%%%%%%%%%%%%%%%%%%%%
\section{The vacuum projection}
 \label{AppendixB}
  %%%%%%%%%%%%%%%%%%%%%%%%%%%%%%%%%%%%%%%%%%%

 In this appendix we prove the formula (\ref{eq:overlap})
\begin{align}
\label{eq:overlaps}
f_{\rK_1|\rK_2}(\xx )=(a_1)^{N_-^{\xx }}(d_2)^{N_+^{\xx }}\,.
\end{align}
where $f_{\rK_1|\rK_2}(\xx )\equiv{_{\rK_1|\rK_2}\langle}\xx
|\!\uparrow^{\LL} \rangle$ is the projection of the SoV basis on the
vacuum state.  The numbers $N_\pm^{\xx }$ are the numbers of pluses
and minuses in the state ${_{\rK_1|\rK_2}\langle}\xx |$ and they are
given in (\ref{nxpm}).  We start with the explicit expression of
$f_{\rK_1|\rK_2}(\xx )$,
\begin{align}
\label{vacover}
f_{\rK_1|\rK_2}(\xx )=\langle\,\uparrow^{\LL} \!|\prod_{k=1}^{\LL}
\left[\frac{A_{\rK_1\rK_2|0}(\theta_k^+)}{Q^+_{\thth
}(\theta_k^+)}\right]^{\frac{1-s_k}{2}}
\rg_{\rK_2}^{-1}|\!\uparrow^{\LL} \,\rangle\,.
\end{align}
It will be useful to write
\begin{align}
\label{grotup}
\rg_{\rK_2}^{-1} \,|\uparrow^{\LL} \,\rangle=d_2^{\,l}\;e^{- \a_2
\,S^-}|\uparrow^{\LL} \,\rangle\,,& \qquad \langle\,\uparrow^{\LL}
\!|=\langle\,\uparrow^{\LL} \!|\,e^{\a_2 \,S^-}\,, \quad \a_2\equiv
c_2/d_2\,.
\end{align}
When all $s_k=+$, the expression (\ref{vacover}) becomes
\begin{align}
f_{\rK_1|\rK_2}(+,\ldots, +)=\langle\,\uparrow^{\LL}
\!|\,\rg_{\rK_2}^{-1}\,|\!\downarrow^{\LL} \,\rangle=(d_2)^{\LL} \,.
\end{align}
When the raising operators $A_{\rK_1\rK_2|0}(\theta_k^+)$ are
present we are going to insert the identity $1=e^{-\a_2\, S^-}e^{+\a_2
\, S^-}$ between consecutive operators, then we compute
\begin{align}
e^{\a_2\, S^-}A_{\rK_1\rK_2|0}(\theta_j^+)\,e^{-\a_2\,
S^-}=e^{\a_2\,S^-}\left( a_{12}\,A^j+b_{12}\, C^j\right)e^{-\a_2\, S^-}\;.
\end{align}
Here and in the following we use for simplicity the notation
$A^j=A(\theta_j+i/2)$, etc.  For this purpose we use the formula
\begin{align}
e^{\alpha \CO_1}\CO_2\, e^{-\alpha
\CO_1}=\CO_2+\alpha[\CO_1,\CO_2]+ \hf{\alpha^2}[\CO_1,[\CO_1,\CO_2]]+\ldots\,,
\end{align}
and the commutators
\begin{align}
\label{commutatorsS}
[S^-,C(u)]=D(u)-A(u)\;, \quad [S^-,A(u)]=B(u)\;, \quad
[S^-,D(u)]=-B(u) \,,
\end{align}
so that
\begin{align}
\label{transAplus}
\tilde A_{\rK_1\rK_2|0}&(\theta_j^+)\equiv e^{\a_2\,
S^-}A_{\rK_1\rK_2|0}(\theta_j^+)\,e^{-\a_2\, S^-}\\[-6mm] \nonumber  \\
\nonumber  &=a_{12}\left(A^j +\a_2\, B^j\right)+b_{12}\(C^j+\a_2\, (D^j-A^j)+\a_2^2\,
B^j\right) \;.
\end{align}
In the next step we use that $D^j\,|\!\uparrow^{\LL}
\,\rangle=C^j\,|\!\uparrow^{\LL} \,\rangle=0$ and
$A^j\,|\!\uparrow^{\LL}
\,\rangle=Q_\theta(\theta_j+i)\,|\!\uparrow^{\LL} \,\rangle$.
Therefore, from the rightmost factor we obtain
\begin{align}
\label{eq:B9}
\frac{\tilde
A_{\rK_1\rK_2|0}(\theta_j^+)}{Q^+_\theta(\theta_j^+)}\,|\!\uparrow^{\LL}
\,\rangle=\left((a_{12}  -\alpha_2b_{12})
-(a_{12} \a_2+b_{12}\,  \a_2^2)\frac{B_j}{Q_\theta(\theta_j+i)}\right)|\!\uparrow^{\LL}
\,\rangle\;.
\end{align}
The first term is what we need to obtain (\ref{eq:overlaps}) since
\begin{align}
a_{12} -\alpha_2b_{12} =
a_1a_2+b_1c_2&-\frac{c_2}{d_2}(a_1b_2+b_1d_2)=\frac{a_1}{d_2}\,.
  \end{align}
The unwanted second term in (\ref{eq:B9}) has to be commuted with the
$A, B, C, D$ terms from the next factors, and finally act on $\langle
\,\uparrow^{\LL} \!|$ on which it will vanish.  The commutators will
mostly give terms which vanish on $|\!\uparrow^{\LL} \,\rangle$, as
one can see from the algebra (the specific coefficients are
irrelevant)
\begin{align}
C^jB^k&\sim B^k C^j+A^kD^j-A^jD^k\;, \\ \nonumber
D^jB^k&\sim B^kD^j+B^jD^k\;,\\ \nonumber
A^jB^k&\sim B^kA^j+B^jA^k\;.
\end{align}
The only non-vanishing terms are those coming from the last line, and
they reproduce $B^j$' s.  Repeating the procedure recursively, and
using that $\langle \,\uparrow^{\LL} \!|\,B^j=0$, one gets the desired
result.  We also need the opposite overlaps $\bar f_{\rK_1|\rK_2}(\xx
)\equiv \langle \,\downarrow^{\LL} \!| \xx \rangle _{\rK_1|\rK_2}$,
expressed as
\begin{align}
\label{vacoverbar}
\bar f_{\rK_{12}}(\xx )=\langle\,\downarrow^{\LL} \!|\,\rg_{\rK_2}
\prod_{k=1}^{\LL} \left[\frac{A_{\rK_1\rK_2|0}(\theta_k^-)}{Q^-_{\thth
}(\theta_k^-)}\right]^{\frac{1+s_k}{2}} |\!\downarrow^{\LL} \,\rangle=
(a_1)^{N_+^{\xx }}(d_2)^{N_-^{\xx }}\,.
\end{align}
The last equality can be proven as above. Again, we write
\begin{align}
\label{grotdown}
|\!\downarrow^{\LL} \,\rangle=\,e^{-\a_2 \,S^-}|\!\downarrow^{\LL}
\,\rangle\,, \quad {\rm and} \quad \langle\,\downarrow^{\LL}
\!|\,\rg_{\rK_2}=(d_2)^{\LL} \,\langle\,\downarrow^{\LL} \!|\,e^{\a_2
\,S^-} \,.
\end{align}
First we consider the case all $s_k=-$, which gives
\begin{align}
\bar f_{\rK_{12}}(-,\ldots,-)=\langle\,\downarrow^{\LL}
\!|\;\rg_{\rK_2}\; |\!\downarrow^{\LL} \,\rangle=(d_2)^{\LL} \,.
\end{align}
When the creation operators are present we use again
\begin{align}
\label{transAminus}
\tilde A_{\rK_{12}|0}(\theta_j^-)&\equiv e^{\a_2\,
S^-}A_{\rK_1\rK_2|0}(\theta_j^-)\,e^{-\a_2\, S^-}=\\[-6mm]
\nonumber  \\ \nonumber
&=a_{12}\left(A_j +\a_2\, B_j\right)+b_{12}\(C_j+\a_2\,
(D_j-A_j)+\a_2^2\, B_j\right)\;.
\end{align}
where now $A_j=A(\theta_j^-)$, {\it etc.} Furthermore, since
$D_j\,|\!\downarrow^{\LL} \,\rangle=B_j\,|\!\downarrow^{\LL}
\,\rangle=0$ and $A_j\,|\!\downarrow^{\LL}
\,\rangle=Q_\theta(\theta_j^-)\,|\!\downarrow^{\LL} \,\rangle$, the
unwanted terms contain $C_j$ at the right and they can be shown to
vanish by using the commutation relations
\begin{align}
B_jC_k&\sim C_kB_j+A_jD_k-A_kD_j\;, \\ \nonumber  D_jC_k&\sim
C_kD_j+C_jD_k\;,\\ \nonumber  A_jC_k&\sim C_kA_j+C_jA_k\;.
\end{align}
 Finally, in order to compute the overlaps between the SoV bases of a
 whole chain and two subchains we need the initial condition
 \begin{align}
\nonumber  \phi(\xx )\equiv{_{\rK_1|0}}\langle+,\cdots,+|&
\otimes{_{0|\rK_2}}\langle-,\cdots,-|\ |\xx
\rangle_{\rK_1|\rK_2}=\langle\,\uparrow^{\LL _{\LL}
}\!|\otimes\langle\,\uparrow^{\LL_1}\!| \ |\xx
\rangle_{\rK_1|\rK_2}=\langle\,\uparrow^{\LL} \!|\xx
\rangle_{\rK_1|\rK_2}\\
&=\langle\,\uparrow^{\LL} \!|\,\rg_{\rK_2} \prod_{k=1}^{\LL}
\left[\frac{A_{\rK_1\rK_2|0}(\theta_k^-)}{Q^-_{\thth
}(\theta_k^-)}\right]^{\frac{1+s_k}{2}} |\downarrow^{\LL} \,\rangle\;.
 \label{incon}
\end{align}
Now we use
\begin{align}
\label{grotupdown}
|\!\downarrow^{\LL} \,\rangle=\,e^{-\b_2 \,S^-}|\!\downarrow^{\LL}
\,\rangle\,, \quad {\rm and} \quad \langle\,\uparrow^{\LL}
\!|\,\rg_{\rK_2}=(b_2)^{\LL} \,\langle\,\downarrow^{\LL} \!|\,e^{\b_2
\,S^-}\, \quad \b_2=a_2/b_2 \,.
\end{align}
Playing the same game as before we obtain that the contribution of
each raising operator amounts to
\begin{align}
\label{cottrans}
e^{\b_2\, S^-}A_{\rK_{12}|0}(\theta_j^-)\, e^{-\b_2\,
S^-}=(a_{12} -\b_2b_{12} )\, A_j+\ldots=-\frac{b_1}{b_2}A_j+\ldots\,,
\end{align}
so that
 \begin{align}
\phi(\xx )=(-b_1)^{N_+^{\xx }}\,(b_2)^{N_-^{\xx }}\,.
\end{align}
It is reassuring to see that none of the various coefficients related
to the SoV basis depend on $c_1$ and $c_2$, which allows one to work
with upper triangular twist matrices.

%%%%%%%%%%%%%%%%%%%%%%%%%%%%%%%%%%%%%%%%%
\section{The splitting function}
\label{AppendixC}
%%%%%%%%%%%%%%%%%%%%%%%%%%%%%%%%%%%%%%%%%

In this subsection, we derive the difference equation for
$\Phi(\yy;\yy_2|\xx )$ and give a solution in terms of
$\Gamma$-functions,.  Let us consider the following quantity
\begin{align}
\langle\yy_1;\yy_2|B_{\rK_1|\rK_2}(u)|\xx \rangle\,,
\end{align}
where we have omitted the indices indicating the twist of the spin
chains.  We first act the operator $B_{\rK_1|\rK_2}(u)$ on the right,
since $|\xx \rangle$ is the eigenstate of the double twisted
$B$-operator, we have
\begin{align}
\langle\yy_1;\yy_2|B_{\rK_1|\rK_2}(u)|\xx \rangle=b_{12}
\prod_{k=1}^{\LL} (u-x_k)\langle\yy_1;\yy_2|\xx \rangle.
\end{align}
On the other hand, we can act the operator $B_{\rK_1|\rK_2}(u)$ on the
left by using the following relation
\begin{align}
B_{\rK_1|\rK_2}(u)=A_{\rK_1|0}(u)B_{0|\rK_2}(u)+B_{\rK_1|0}(u)D_{0|\rK_2}(u).
\end{align}
By taking $u={y}_{1,i}$ and $u={y}_{2,j}$, we can derive two sets of
difference equations
\begin{align}
\label{diffeq}
&b_{12} Q_\xx(y_{1, j}) \Phi(\yy_1;\yy_2|\xx )=b_2\, Q_{\yy_2} (y_{1,
j} )\, Q_{\thth _{1} }^+(y_{1,
j})\Phi(\cdots,y_{1,j}-i,\cdots;\yy_2|\xx )\\\nonumber   &b_{12}\,
Q_\xx(y_{2,j} ) \Phi(\yy_1 ;\yy_2|\xx )=b_1 Q_{\yy_1} (y_{2,j}) \,
Q_{\thth _2 }^-(y_{2,j})\Phi(\yy_1 ;\cdots,y_{2,j}+i,\cdots|\xx ).
\end{align}
Before solving these equations, let us note that the above two sets of
difference equations (\ref{diffeq}) do not shift $x_k$.  This implies
that when we solve the above equations, we treat $x_k$ as constant and
the solution of these equations can only fix $\Phi(\yy_1 ;\yy_2|\xx )$
up to an initial condition $\phi(\xx )$ defined in equation
(\ref{incon}) and computed in appendix \ref{AppendixB}
\begin{align}
\phi(\xx )=(-b_1)^{N_+^{\xx }}\,(b_2)^{N_-^{\xx }}\,.
\end{align}
The strategy of solving equations of type (\ref{diffeq}) is to assume
that the solution takes factorized form and decompose the equations
into simpler difference equations which can be solved
straightforwardly in terms of $\Gamma$-functions.  Let us start by
assuming that the solution of the difference equation takes the
following form
\begin{align}
\phi_s(\yy_1 ;\yy_2|\xx )=\rho(\yy_1 ;\yy_2|\xx )\, \varphi(\yy_1
;\yy_2|\xx )
\end{align}
where $\rho(\yy_1 ;\yy_2|\xx )$ satisfies the equations
\begin{align}
&\rho(\yy_1 ;\yy_2|\xx )=\frac{b_2}{b_{12}} \, Q_{\yy_2}(y_{1, j})\,
\rho(\cdots,y_{1,j}-i,\cdots;\yy_2|\xx )\\\nonumber &\rho(\yy_1
;\yy_2|\xx )=\frac{b_1}{b_{12}} \, Q_{\yy_{1}}(y_{2, j})\, \rho(\yy_1
;\cdots,y_{2,j}+i,\cdots|\xx ).
\end{align}
The function $\varphi(\yy_1 ;\yy_2|\xx )$ has to take care of the
remaining factors in the difference equations (\ref{diffeq}) and
satisfies the following equations
\begin{align}
\label{phi}
& Q_\xx(y_{1, j}) \, \varphi(\yy_1 ;\yy_2|\xx )=Q_{\thth _{1}
}^+(y_{1,j})\varphi(\cdots,y_{1,j}-i,\cdots;\yy_2|\xx )\\\nonumber &
Q_{\xx}(y_{2, j})\, \varphi(\yy_1 ;\yy_2|\xx )=Q_{\thth _2
}^-(y_{2,j})\varphi(\yy_1 ;\cdots,y_{2,j}+i,\cdots|\xx ).
\end{align}
It turns out the solution of equations in (\ref{phi}) has the
following factorized form
\begin{align}
\varphi(\yy_1 ;\yy_2|\xx )=\prod_{j=1}^{\LL _{1} }\varphi_{1} (y_{1,j}|\xx )
\prod_{j=1}^{\LL_2}\varphi_2 (y_{2,j}|\xx )
\end{align}
where $\varphi_{1} (y|\xx )$ and $\varphi_2 (z|\xx )$ satisfy the
following equations
\begin{align}
& Q_{\xx}(y) \varphi_1(y|\xx )=Q_{\thth _{1} }^+(y)\varphi_{1}
(y-i|\xx ),\\\nonumber & Q_\xx(y)\, \varphi_2(z|\xx )=Q_{\thth _2
}^-(z)\varphi_2 (y+i|\xx ).
\end{align}
The solution is in the end  given by (\ref{eq:overlap}):
$\Phi(\yy_1;\yy_2|\xx )= \texttt{twist}\times \texttt{Gamma}$, with
\begin{align}
\nonumber
\texttt{twist}&= \(i{b_{12}/ b_1}\)^{N_+^{\yy_2}} \(-i {b_{12}
/b_2}\)^{N_-^{\yy_1}} \,\left(-b_1\right)^{N_+^{\xx
}}\left(b_2\right)^{N_-^{\xx }},
 \nonumber    \\\nonumber
  \\
\nonumber \texttt{Gamma}&=\frac{\Gamma\left(i(\thth _{1} ^+-\thth _2
^-) \right)}{\Gamma\left(i(\yy_1 -\yy_2)\right)}
\frac{\Gamma\left(1-i(\yy_1 -\thth _{1}
^-)\right)}{\Gamma\left(1-i(\yy_2-\thth _2 ^-)\right)}
\frac{\Gamma\left(1+i(\xx- \thth _{1} ^+)\right)}{\Gamma\left(1+i(\xx-
\yy_1 )\right)}\, \frac{\Gamma\left(1-i( \xx- \thth _2 ^-
)\right)}{\Gamma\left(1-i(\xx-\yy_2 )\right)}.
\end{align}

\bibliographystyle{JHEP}
\providecommand{\href}[2]{#2}\begingroup\raggedright\endgroup

\end{document}